\documentclass[natbib,graybox]{svmult}

\usepackage{mathptmx}       
\usepackage{helvet}         
\usepackage{courier}        
\usepackage{type1cm}        
\usepackage{makeidx}         
\usepackage{graphicx}        
\usepackage{multicol}        
\usepackage[bottom]{footmisc}

\makeindex             

\begin{document}

\title*{The Milky Way Bulge: Observed properties and a comparison to external galaxies}
\author{Oscar A. Gonzalez and Dimitri A. Gadotti}
\institute{Oscar A. Gonzalez \at European Southern Observatory, Alonso de Cordova 3107, Vitacura, Santiago, Chile, \email{ogonzale@eso.org}
\and Dimitri A. Gadotti \at European Southern Observatory, Alonso de Cordova 3107, Vitacura, Santiago, Chile, \email{dgadotti@eso.org}}
%
%
\maketitle

\abstract*{}

\abstract{The Milky Way bulge offers a unique opportunity to investigate in detail the role that different processes such as dynamical instabilities, hierarchical merging, and dissipational collapse may have played in the history of the Galaxy formation and evolution based on its resolved stellar population properties. Large observation programmes and surveys of the bulge are providing for the first time a look into the global view of the Milky Way bulge that can be compared with the bulges of other galaxies, and be used as a template for detailed comparison with models. The Milky Way has been shown to have a box/peanut (B/P) bulge and recent evidence seems to suggest the presence of an additional spheroidal component. In this review we summarise the global chemical abundances, kinematics and structural
properties that allow us to disentangle these multiple components and provide constraints to understand their origin. The investigation of both detailed and global properties of the bulge now provide us with the opportunity to characterise the bulge as observed in models, and to place the mixed component bulge scenario in the general context of external galaxies. When writing this review, we considered the perspectives of researchers working with the Milky Way and researchers working with external galaxies. It is an attempt to approach both communities for a fruitful exchange of ideas.}

\section{Introduction}
\label{sec:1}

What is the origin of the bulge of the Milky Way? The answer to this question is a crucial step towards identifying the history of events that took place during the formation and evolution of the Galaxy as a whole. As a matter of fact, the answer to this question has changed dramatically, since the early times of Galactic bulge archaeology until today, between the two main ideas behind bulge formation: the merger-driven bulge scenario, where a bulge is formed violently and quickly during the early stages of the Galaxy dominated by the gravitational collapse or hierarchical merging of sub-clumps of dark matter carrying baryons and gas \citep[e.g.][]{abadi+03, elmegreen+99}, and the secular evolution scenario where the bulge structure is naturally born from the dynamical evolution of the stellar Galactic disc \citep{ComSan81, raha+91, norman+96, Ath05b}. Until this day, we cannot say with absolute certainty which one of these scenarios, or if perhaps both, played a major role during the formation of the bulge of our Galaxy\footnote{To be complete, we must mention the pioneering work of \citet{EggLynSan62}, who suggested that the first stage in the formation of the Milky Way was a fast ($<10^8$ yr) monolithic collapse of a single massive gas cloud, which could have formed both the Galaxy stellar halo and the bulge. This scenario was later replaced with the merger-driven scenario mostly due to the widely varying ages of different components of the Galaxy and the hierarchical nature of LCDM theory. We refer the reader to Brooks \& Christensen (this volume) for a review on merger-driven bulge formation. In the last decade or two, the secular evolution scenario has slowly, but unequivocally, gained terrain over the merger-driven scenario. Another bulge building scenario we do not discuss here has recently been put forward by e.g. \citet[][-- see review by Gadotti 2012]{elmegreen+08}. In this scenario, bulges form by the coalescence of giant clumps in primordial discs. This scenario can explain the formation of spheroids but does not account for box/peanuts.}.

The reason for this long term debate might lie, ironically, in our greatest strength. The Galactic bulge allows us to investigate its properties by taking advantage of the fully resolved stellar populations - a unique strength that can be understood as the ability to see its properties in a unique level of detail with respect to what we can learn from the observation of external bulges. However, such an advantage also means that nearly 500 sq. deg. of sky must be homogeneously covered in order to obtain the most global picture of the Bulge. To obtain a general characterisation of these properties is a crucial step in order to answer the question of the origin of our Galactic bulge.

As a consequence, observational efforts during the last decade have been focused on solving this limitation. As a result, our knowledge regarding the global properties of the Galactic bulge has increased considerably thanks to the advent of dedicated spectroscopic and photometric surveys. We are currently witnessing a revolution in the field of Milky Way bulge research that will also find its place within the bulges of other disc galaxies. It is thus a moment in which the communities of Galactic and Extragalactic research are approaching each other. From this, one cannot expect anything but a fruitful exchange of ideas that will certainly push both fields forward. However, it is not straightforward for members of each community to study the other field, for at least two reasons. The first obstacle is the vast amount of work, rich in details, that one has to become familiarised with. The second obstacle is the jargon employed independently by each group which hampers understanding. This review is a modest first attempt to overcome these obstacles.

\section{The structure of the Milky Way bulge}
\label{sec:2}

The most basic definition of a galactic bulge is that of an over-density that swells up from the plane of the disc. The idea of this natural conception originated from the observations of other disc galaxies, in particular the bulges of edge-on spirals, which allow us to compare them more easily with our own in a purely morphological way.

Within spiral galaxies, we could in principle distinguish between classical bulges and those bulges which are formed via secular evolution, based almost solely on their morphological signatures. Clearly, a proper characterisation of the structural properties of the Milky Way bulge would provide us with a valuable set of constraints needed to find its place in the more general scheme of external bulges and, as extensively shown by galaxy formation models, to connect these constraints to the different mechanisms of origin.

After it was first postulated by \citet{deVaucouleurs+64}, followed by \citet{sinha+79} and \citet{liszt+80} among others, \citet{blitz+91} predicted the presence of the bar-like structure for the inner regions of the Galaxy based on infrared observations. Later on, continuing with the exploitation of infrared imaging in order to overcome the strong dust obscuration towards the inner galaxy, the COBE/Diffuse Infrared Background Experiment \citep{smith-price-baker+04} data was used by \citet{weiland+94} to unambiguously establish the presence of the bar. Soon after, the COBE data further revealed the global B/P morphology of the Milky Way bulge \citep{dwek+95}.

An important number of Bulge structural studies have been based on the stellar counts of red-clump stars, which are the metal-rich counterpart of the well known globular cluster horizontal-branch stars. The absolute magnitudes of red-clump stars, are found to have little dependence on age and metallicity, making them one of the most powerful tools for deriving distances towards the bulge and therefore tracing its global morphology. This method is based on the construction of the luminosity function of the Bulge towards a given line of sight where the red-clump feature can be easily identified and fitted with a Gaussian distribution to obtain the mean red-clump magnitude \citep{stanek+94}. \citet{zoccali+10} and \citet{mcwilliam-fulbright-rich+10} presented the discovery of a split within the red-clump when investigating the luminosity function of the Bulge at latitudes $\rm |b| > 5$, along the minor axis. Soon enough, \citet{mcwilliam-zoccali+10} and \citet{nataf+10} provided a wider mapping of this split red-clump in the color magnitude diagram, providing substantial evidence for the bright and faint red-clumps to be the consequence of having two over-densities of stars located at different distances, namely the two southern arms of an X-shaped structure both crossing the lines of sight. Detailed three-dimensional maps were later constructed by \citet{saito+11}, based on the use of red-clump stars observed in the near-IR survey 2MASS, which confirmed the suggestion of \citet{mcwilliam-zoccali+10} that the Bulge is in fact X-shaped due to the prominent vertices of the B/P. \citet{wegg-gerhard+13} modelled the distribution of red-clump stars, observed in Vista Variables in the Via Lactea (VVV) ESO public survey, providing the first complete mapping of the X-shaped Bulge. These X-shaped bulges are commonly observed in external edge-on galaxies and belong to the case of a pronounced B/P structure, that is simply the inner regions of the bar that grow out of the plane of the disc. 

Currently, the axial ratios of the bar are constrained to be about 1:0.4:0.3 with a bar size of about 3.1$-$3.5 kpc diameter and the near end of the bar pointing towards positive Galactic longitudes. Until recently, the position angle of the bar was constrained to a relatively large range of values between $\sim$20$-$40 degrees with respect to the Sun–centre line of sight. The uncertainty in the bar position angle is likely to be a consequence of measurements done across different latitudes in each study, thus finding a different position angle when looking at different distances from the Galactic plane. As a matter of fact, specific evidence for a longer flatter component of the bar, referred to as the Galactic long bar, has been presented in the literature based on near-IR star counts near the Galactic plane \citep{blitz+91, stanek+94, dwek+95, binney+97, bissantz-gerhard+02, benjamin+05, babusiaux+05, rattenbury+07, cao+13}. This long bar is found to have an axis length of 4$-$4.5 kpc and ratios of 1:0.15:0.03. The position angle of this longer component has been constrained to $\sim$45 degrees in such studies \citep[e.g.][]{lopez-corredoira+07,cabrera-lavers+07,hammersley+00,churchwell+09,amores+13}. On the other hand, the recent model of the global distribution of red-clump stars from \citet{wegg-gerhard+13} provided a precise measurement for the B/P Bulge position angle of $27\pm 2$ degrees, in agreement with the studies done at larger distances from the Galactic plane. The nature of the long bar has been debated extensively in the literature. Recently, \citet{garzon+14} provided a theory where two co-existing bars, the long bar restricted to the plane latitudes and the B/P thick bar, could be present in the inner Galaxy. However, model observations of barred galaxies led \citet{martinez-valpuesta+11}, \citet{romero-gomez+11}, and \citet{athanassoula+12} to strongly argue that the apparent long bar is an artefact associated with leading spiral features at the end of the shorter primary bar (the B/P Bulge). Furthermore, the co-existence of such independently large scale structures has not been seen in external galaxies. For this reason, the observed properties attributed to different bars in the Galaxy are more likely corresponding to a unique B/P bulge and bar structure formed by the buckling instability process. The long bar would then be explained by the interaction of the outer bar, with the adjacent spiral arm near the plane which produced leading ends that ultimately results in the measurement of a larger position angle \citep{martinez-valpuesta+11}.

In the innermost regions ($\rm l < 4$, $\rm b < 2$) the bar has been found to change its apparent inclination with respect to the line of sight which has been interpreted as evidence for a possible distinct smaller bar, referred to as a nuclear bar. However, models of a single bar (meaning those that do not include a distinct nuclear bar) have also shown such a change in orientation in the inner regions, most likely due to the presence of a more axisymmetric concentration of stars in the central regions \citep{gerhard+12}.

Red-clump stars, although an excellent distance indicator for the bulge mean population, still suffer from the usual complications when looking towards the inner Galaxy such as disc contamination, extinction, and, to a minor extent, the effects of stellar populations. Furthermore, red-clump stars will map the distribution of the variety of stellar ages of the Bulge population, which is not defined beforehand. This uncertainties can be statistically handled when constructing and analysing the Bulge luminosity function, however their impact on the results will depend on the level of knowledge of the properties of each analysed field. For this reason, Variable stars, specifically RR-Lyrae, have recently provided a new perspective for the Bulge structural properties. Their well defined period-luminosity relation in the near-IR helps to overcome the effects of dust extinction in the inner Galaxy and they are well spatially distributed across the entire bulge. The period-luminosity relation makes RR-Lyrae an exquisitely accurate distance indicator that unequivocally traces the oldest Galactic population.  

\begin{figure}[]
\centering
\sidecaption
\includegraphics[angle=0, scale=1.3,trim=0cm 0cm 0cm 0cm,clip=true]{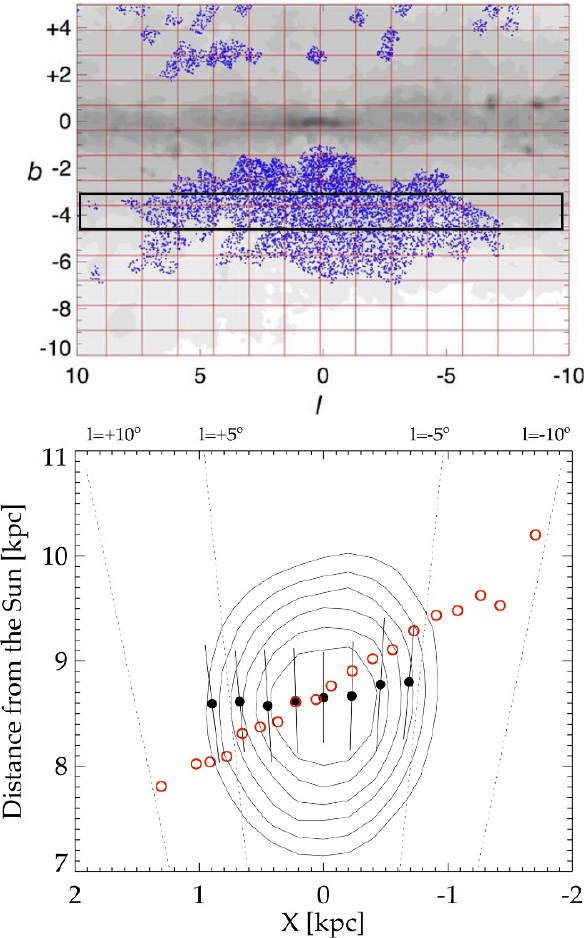}
%
%
\caption{Upper panel: Spatial distribution in Galactic coordinates for the 7663 OGLE-III RRab stars in the bulge area of the VVV Survey, from \citet{dekany+13}. The grey scale background shows the interstellar extinction map of \citet{schlegel+98}. The black rectangle denotes the region at $\rm b=-4$ for which the mean distance of RR Lyrae and RC stars are compared in the lower panel. The lower panel shows the projected mean distances of RR Lyrae in black filled circles and of the red-clump stars as red open circles. Isodensity contours for the projected distance distribution of the RR Lyrae sample in the analysed latitude range are also shown. Mean distances of the red clump stars are from the mean magnitudes obtained in \citet{gonzalez+12} and calculated adopting an absolute magnitude of $\rm M_{Ks,RC}=1.71$ mag. Distances for the sample of RR Lyrae have been presented in \citet{dekany+13} and were used here to derive the projected mean distance and 1$\sigma$ width (black solid lines) to each line of sight. [{\it Upper panel adapted from \citet{dekany+13}.}]}
\label{rrlyrae_dist}       
\end{figure}

Surprisingly, given the vast amount of different tracers that have confirmed a dominant barred structure, RR Lyrae have shown a remarkably different spatial distribution compared to, for example, red-clump stars. While red-clump stars trace the position angle of the bar at all latitudes, a direct comparison with the RR Lyrae distance distribution provided by \citet{dekany+13} strongly suggests a different morphology for the oldest population in the inner Galaxy. Unlike the red-clump stars, the RR Lyrae stars show a more spheroidal, centrally concentrated distribution. This structural component, populated by stars with ages larger than 10 Gyr, seems to be overlying with the B/P bulge. Figure~\ref{rrlyrae_dist} shows a comparison of the projected mean distances obtained from RR Lyrae and those from the mean magnitude distribution of red-clump stars. The figure illustrates the structural difference between the components traced by both distance tracers, with only red-clumps stars following the position angle of the bar. This result presented in \citet{dekany+13} is perhaps the first purely morphological evidence suggesting a composite Bulge nature, with two different stellar populations overlapping in the inner Galaxy. 

%

The presence of more than one age/metallicity distribution within a B/P bulge has already been seen in dissipative collapse models \citep[e.g.][]{SamGer03} and also in bulges from cosmological galaxy formation simulations \citep[e.g.][]{ObrDomBro13}. However, one must be very careful when further linking the different spatial distributions seen in the Galactic bulge with a distinct origin process, namely having a classical bulge and a secularly evolved B/P bulge that originated from the disc. Recently, \citet{ness+14} gives caution to the fact that different spatial distributions and mean stellar ages can be found in pure B/P bulges without the need of a merger-origin structure to be present, as seen in an N-body + smoothed particle hydrodynamics simulation of a disc galaxy. Certainly, the fine details of the shape traced by RR-Lyrae will be achieved when the complete sample of RR-Lyrae from the VVV survey is available. The mapping of a wider area of the Bulge and the larger sample of sources available in each line of sight will allow for the calculation of precise mean distances with high spatial resolution. 

\section{The age of the Milky Way bulge}
\label{subsec:2}

Right from the very early discovery of RR Lyrae towards the centre of the Galaxy \citep{baade+51}, perhaps the most common statement found in the literature addressing the Galactic bulge stellar populations is: \textit{The Milky Way bulge stellar population is predominantly old}. The reasons for such a statement are indeed very well funded and are described in this section.

Without a doubt, deep photometric observations towards low extinction regions of the Bulge provided a defining view on the age of its stellar population thanks to the possibility of constructing colour-magnitude diagrams that reached the turn-off position - a useful indicator for the mean age of a given stellar population. However, the effects of differential reddening and, in particular, the uncertainty on the distance modulus of the stars towards a given line of sight does not allow the derivation of an absolute age estimation for the Bulge using the turn-off technique. To overcome these issues, \citet{ortolani+95} adopted a differential method, based on the position of red-clump stars in the luminosity function. The very small dependence on age and metallicity for the mean magnitude of the red-clump allowed \citet{ortolani+95} to match its position in the luminosity function to that of an old stellar population, the globular cluster NGC 6528, finding a remarkable agreement between the relative positions of the turn-off. This result provided strong support for the view of the Galactic Bulge being formed in a time scale shorter than 1 Gyr, thus making it as old as the globular cluster population.

Dedicated photometric studies spread across other regions of the Bulge, and based on similar techniques further strengthen the conclusion of a stellar population with a mean age of $\sim10$ Gyr, particularly setting a lower limit on ages higher than 5 Gyr \citep{zoccali+03,  valenti+13, clarkson+11}. The major issue with these kind of studies is how to deal with the contamination of the foreground disc, in particular with the main sequence of the disc which lies right on top of the Bulge turn-off. Statistical decontamination methods, for example using disc control fields \citep{zoccali+03, valenti+13}, have been used to eliminate foreground stars to some extent. However,  the contamination of foreground stars coupled with uncertainties in differential reddening and metallicity distribution effects lead to uncertainties on age determination via the analysis of color-magnitude diagrams that remain of the order of 2 Gyr. 

%
\begin{figure}[]
\sidecaption
\centering
\includegraphics[angle=0, scale=.40,trim=0cm 0cm 0cm 0cm,clip=true]{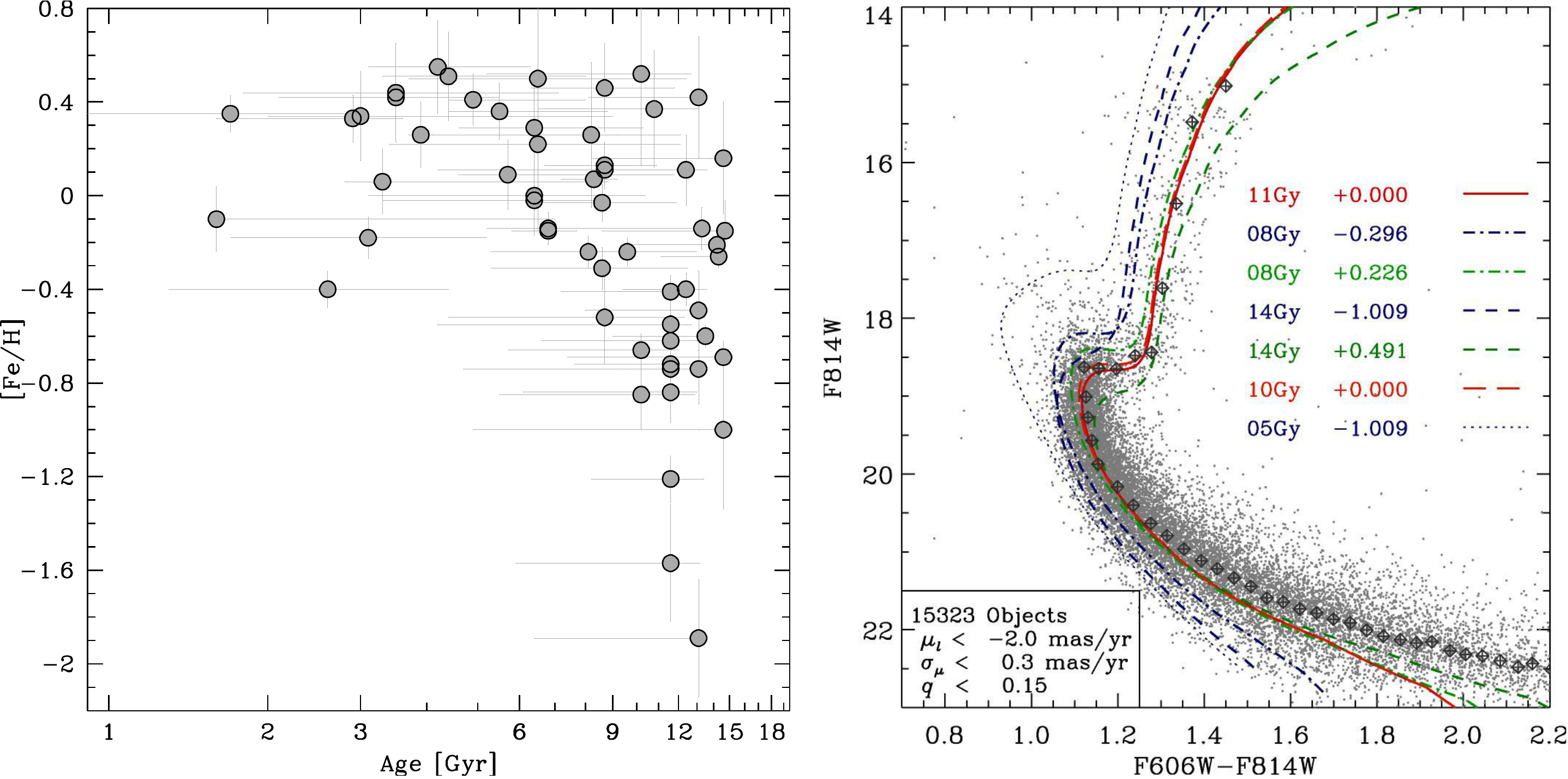}
%
%
\caption{Left panel: The metallicity of the microlensed dwarfs of the Bulge as a function of their ages taken from \citet{bensby+13}. Right panel: The color magnitude diagram for proper motion-selected Bulge objects from \citet{clarkson+08},  using similar mean proper motion criteria to \citet{kuijken-rich+02} but with a 6$\sigma$ detection requirement imposed. A set of isochrones with different metallicities and ages is overplotted to the color magnitude diagram. An alpha-enhanced, solar-metallicity isochrone at 11 Gyr represents the median sequence well above the turn-off. Also shown in the figure are sequences at metallicity [Fe/H]=(-1.009, -0.226, +0.491) and ages (8, 10, 14) Gyr.  [{\it Left panel adapted from Fig. 15 in \citet{bensby+13}, \copyright ESO reproduced with permission. Right panel adapted from Fig. 20 in \citet{clarkson+08}, \copyright AAS reproduced with permission. }]}
\label{bulge_age}       
\end{figure}

Recently, \citet{clarkson+08} provided what is perhaps the cleanest, and thus most accurate, age determination of those studies based on the analysis of a colour-magnitude diagram (Fig.~\ref{bulge_age}). \citet{clarkson+08} were able to decontaminate the turn-off position of the Bulge population from that of the nearby disc by using a mean proper motion criteria as the one shown by \citet{kuijken-rich+02} with a set of ACS WFC on HST observations taken over 123-orbit HST integrations in the SWEEPS field (l,b)=(1.25, -2.65). The study of \citet{clarkson+11} later concluded that a fraction of only up to a 3.5\% of the Bulge stars can be younger than 5 Gyr. Therefore, this study is in agreement with most of other studies where the Bulge population is found to be dominantly old. 

A large percentage of young stars is certainly not expected in a bulge that originates purely from early dissipation or merging processes. These events would take place at early times in the evolution of the galaxy and would occur rapidly. Although some intermediate-age stars could later be added to the bulge by diffusion from the inner disc, the bulk of the bulge stellar population would be old.  On the other hand, a significant population of young stars could be found $-$ and might be actually expected $-$ in bulges formed via disc instabilities. So if the Milky Way bulge is certainly dominated by a component formed via disc instability, as seen by its structural properties, then where are those young stars? It is only very recently that a possible answer to this question has been brought to the table, and it was using a completely different approach: the high-resolution spectroscopy of Bulge microlensed stars. The microlensing event, which produces the brightness magnification of a dwarf star in the bulge due to the passing of a foreground lens star, provides a unique opportunity to obtain high resolution spectra of this otherwise unreachable target. \citet{bensby+13} have collected enough micro-lensed dwarfs to investigate the overall metallicity distribution and also their age. While the metallicity distribution of the microlensed dwarfs has been found to be in good agreement with that of the bulge giant stars, their age distribution presented a significant number of young stars: nearly $22\%$ of the micro-lensed dwarfs were found to be younger than 5 Gyr. As the number of analysed micro-lensed stars increases, these findings will be further confirmed or disproved with better statistics in order to refine the actual percentages of young and old stars found in the bulge. Particularly, notice the disagreement between the $22\%$ of stars younger than 5 Gyr found in the microlensing sample and the corresponding $3.5\%$ fraction of young stars given in the work of \citet{clarkson+11}. 

\section{The chemical abundances of the Milky Way bulge}
\subsection{The metallicity distribution}

The characterisation of the chemical abundance of Bulge stars is perhaps the field that has evolved most quickly thanks to the advent of multi-object spectrographs in large telescopes. Different surveys have $-$ and still are $-$ pointing towards different regions of the bulge and collecting samples of thousands of stars. The argument for such surveys is clear: it became evident that a few low extinction regions were no longer representative of the global chemical abundance patterns of the Bulge. First attempts to derive the metallicity distribution of the Bulge based on low resolution spectra  \citep[e.g.][]{sadler+96, ramirez+00}, together with a small number of available spectra obtained with high resolution \citep{McWRic94, fulbright-mcwilliam-rich+06}, had shown from the start that the mean population of the bulge was overall metal-rich. The shape of the metallicity distribution was, on the other hand, less clear. \citet{zoccali+03} used a set of observations using WFI to obtain a photometric metallicity distribution based on the colour of red giant stars. They found a rather broad metallicity distribution, in good agreement with that derived from spectroscopy in the same field but with a larger statistical sample, with [Fe/H] values ranging from -1.0 to 0.4 and which peak at solar metallicity. 

With a well-characterised metallicity distribution in Baade's window, it was time to answer the following question: how spatially uniform were these properties? \citet{minniti+95} had already discussed the possibility for a metallicity gradient in the Bulge, impressively enough based on low resolution spectra of less than a hundred giant stars. It then became the era of multi-object spectroscopy where hundreds of stars could be observed in one single shot. Using FLAMES on the Very Large Telescope, \citet{zoccali+08} derived the metallicity distribution for different fields along the Bulge minor axis at different latitudes ($b=-4$, $-6$, and $-12$). They found a clear metallicity gradient of $\sim$0.6 dex/kpc, with mean metallicities varying from -0.4 at the largest latitudes and up to solar metallicity at b=-4. This gradient has since then been confirmed thanks to several subsequent observations across different regions and the variation in the metallicity distributions have been further characterised \citep{johnson+11, uttenthaler+12, johnson+13, ness-abu+13, rojas-arriagada+14}.

%
\begin{figure}[]
\sidecaption
\centering
\includegraphics[angle=0, scale=1.42,trim=0cm 0cm 0cm 0cm,clip=true]{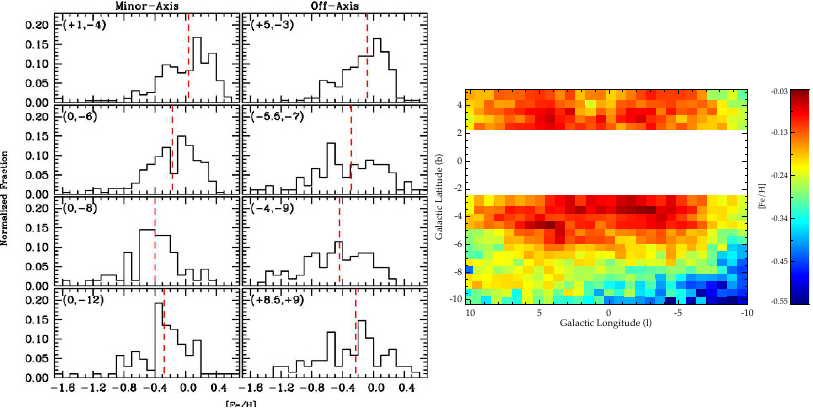}
%
%
\caption{Left panel: Metallicity distributions for a compilation of studies taken from \citet{johnson+13}. The $\rm (l, b) = (+1, -4), (0, -6)$ and $\rm(0, -12)$ fields are from \citet{zoccali+08}, the $\rm (0, -8)$ field is from \citet{johnson+11}, the $\rm (+5, -3)$ field is from \citet{gonzalez+11}, and the $\rm (-5.5, -7), (-4, -9)$, and $\rm (+8.5, +9)$ fields are from \citet{johnson+13}. Right panel: Map of the mean photometric metallicities of the bulge constructed with the VVV survey data from \citet{gonzalez+13}. [{\it Left panel adapted from Fig. 8 in \citet{johnson+13}, \copyright AAS reproduced with permission. Right panel adapted from Fig. 2 in \citet{gonzalez+13}, \copyright ESO reproduced with permission.}]}
\label{met_grad}       
\end{figure}

\citet{gonzalez+13} recently complemented these results by presenting a photometric metallicity map, constructed with the same technique used by \citet{zoccali+08} but based on the Vista Variables in the Via Lactea (VVV) ESO public survey, for almost the entire Bulge region providing the global picture of the Bulge metallicity gradient. The metallicity gradient is therefore strongly established by an increasing number of spectroscopic studies obtained with different techniques and stellar samples (see Fig.\ref{met_grad} for a compilation of the latest results). However, the metallicity distributions obtained in the innermost regions of the Bulge ($|b|<4$) based solely on high-resolution, near-infrared spectroscopy have provided evidence for the flattening-out of the gradient in the inner 700 pc \citep{ramirez+00, rich-origlia-valenti+12}. 

What is the implication of finding such a metallicity gradient in the Bulge? At first, similarly to the domination of old ages found in Bulge stars, the metallicity gradient was interpreted as direct evidence for a bulge formed as a classical bulge via mergers in the early stages of the galaxy, similarly to elliptical galaxies. It was also interpreted as evidence against the secular evolution scenario, since it was thought that bars would mix the stellar orbits well enough to erase any existing vertical gradient. Models of bar formation in disc galaxies, however, proved otherwise, showing that a bar might produce a gradient similar to the one seen in the Milky Way depending on the original disc radial gradients \citep{martinez-valpuesta+13}, vertical gradients \citep{bekki+11}, or both \citep{dimatteo+14}.

\begin{figure}[]
\sidecaption
\includegraphics[angle=0, scale=.55,trim=0cm 0.0cm 0.0cm 0cm,clip=true]{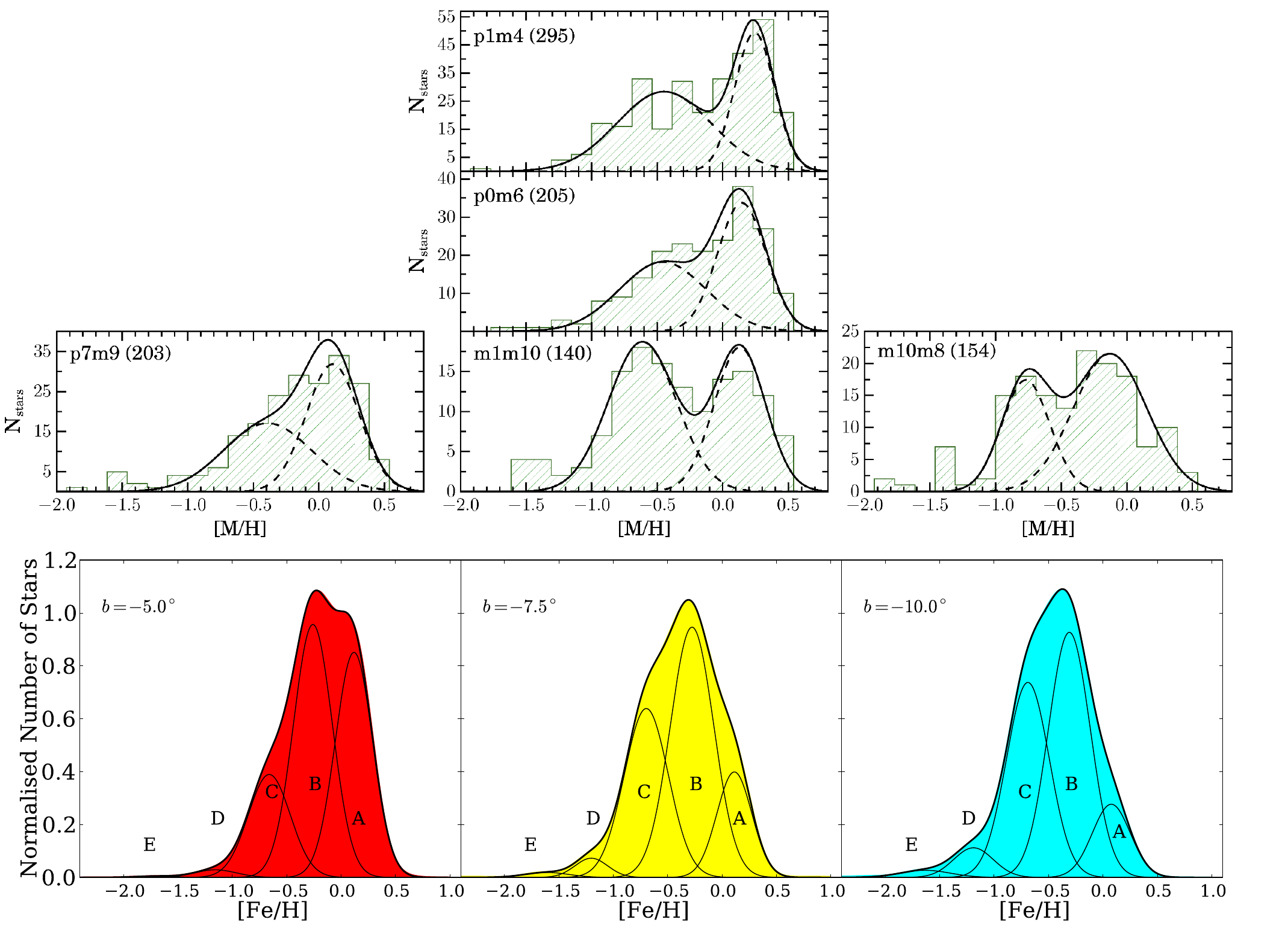}
%
%
\caption{Upper panels: Metallicity distributions for the five bulge fields from the ESO Gaia survey, presented in \citet{rojas-arriagada+14}. Black dashed and solid lines show the components identified in each field as dashed Gaussian functions, with the sum of them shown as a solid line. The three minor-axis fields located at (+1,-4), (0,-6), and (-1,-10) are shown in the central panels, and the lateral fields (+7,-9) and (-10,-8) at the left and right. Lower panels: Metallicity distributions from the ARGOS survey from \citet{ness-kine+13}. From left to right at b = $-5$, $-7.5$ and $-10$, for  l = $\pm15$. The different contribution of the adopted Gaussian components are marked in each field, with the three main components being A, B and C.  [{\it Upper panel adapted from Fig. 6 in \citet{rojas-arriagada+14}, \copyright ESO reproduced with permission. Bottom panel adapted from Fig. 1 in \citet{ness-kine+13} }]}
\label{argos_mdf}       
\end{figure}

However, the existence of metallicity gradients has also been interpreted as a consequence of having two or more underlying components each one with a characteristic metallicity distribution. This mixing of components would naturally produce a variation on the mean metallicity according to the bulge region which is being studied. Evidence for such a multiple component scenario has been suggested based on a bimodal metallicity distribution of red clump stars in Baade's window by \citet{hill+11} and has been also suggested from a similar bi-modality seen in the metallicity distribution of microlensed dwarfs. Both of this distributions show a metal-poor and a metal-rich peaks located approximately at $\rm [Fe/H]\sim-0.3$ and $\rm [Fe/H]\sim+0.3$, respectively. Recently, the same bi-modality has also been found in the metallicity distributions based on the Gaia-ESO survey observations of the Bulge \citep{rojas-arriagada+14}. By producing a Gaussian decomposition of the metallicity distribution functions, \citet{rojas-arriagada+14} showed a clear bi-modality in all the analysed fields with relative sizes of components depending of the specific position on the sky. This change in the relative sizes of each component can be clearly seen in Fig.~\ref{argos_mdf}.

%

On the other hand, the ARGOS Galactic bulge survey \citep{freeman-argos+13}, which consists of the largest sample of homogeneously analysed RC stars, constructed large-number statistics metallicity distributions at different latitude stripes, using a total of more than 10,500 stars located within a galacto-centric radius of 3.5 kpc. The overall metallicity distribution of the ARGOS survey was interpreted in \citet{ness-abu+13}, as being composed of five Gaussian components. Each of these components would be sampling a different stellar population and thus any changes in their relative contribution fraction as a function of latitude could be the origin of the observed mean metallicity gradient seen in the Bulge. The metallicity distributions for stars within Galactic longitudes $l=\pm15$ and latitudes $b=-5$, $-7.5$, and $-10$ from the ARGOS survey \citet{ness-abu+13} are also shown in Fig.~\ref{argos_mdf}. This effect has lead the ARGOS survey to suggest the three main components of the metallicity distribution to be associated with the metal rich B/P bulge (mean [Fe/H]  $\sim+$0.15), the thick B/P bulge (mean [Fe/H]$\sim-$0.25) and the inner thick disc (mean [Fe/H]$\sim-$0.70). 

Although the results from \citet{ness-abu+13} are based on the identification of several components, it can be safely understood that most studies converge into a similar conclusion, the bulge metallicity distribution is the result of not a single but of a mixture of populations with at least two main components, one metal-poor and the other metal-rich. Certainly, this is everything that can be concluded from the metallicities alone and one has to be extremely cautious when attempting to link these different components with a given bulge formation scenario. All evidence needs to be considered when interpreting results in terms of such scenarios, i.e. such an assessment must also include the kinematics, spatial distribution and ages of the stars. 

\subsection{The Bulge alpha-element abundances}

The $\alpha$-element abundances can provide us with further constraints for the origin of the Bulge stellar populations, specifically with respect to its formation time-scale. \citet{tinsley+79} suggested that the ratio of [$\alpha$/Fe] compared to [Fe/H] is a function of the time delay between SNe II, which produce both $\alpha$- and iron-peak elements \citep[e.g.][]{woosley+95}, and SNe Ia, which yield mostly iron-peak with little $\alpha$-element production \citep[e.g.][]{nomoto+84}. Therefore, only after sufficient time has passed for the SNe Ia events occur, the [$\alpha$/Fe] ratio will decline from the SNe II value. Clearly, the critical ingredient on this relation is the SNe Ia delay time, for which different production channels might be present. 

In the Bulge, the $\alpha$-element abundances of Bulge stars with [Fe/H]$<$-0.3 have been found to be enhanced over iron by [$\alpha$/Fe]$\sim$+0.3 dex \citep{McWRic94, rich-origlia+05, cunha-smith+06, fulbright-mcwilliam-rich+07, lecureur+07, rich-origlia-valenti+07} calling for a fast formation scenario, while metal-rich stars [Fe/H]$>$-0.3 showed a decrease in [$\alpha$/Fe] reaching solar values for metallicities larger than Solar. However, is important to note that not all elements were found to follow the same yield trends. 

A relative approach has been commonly adopted instead of an absolute interpretation of the Bulge $\alpha$-element ratio. The direct comparison of [$\alpha$/Fe] values in Bulge stars against those of other galactic components then provides a relative time constraint on the bulge formation. \citet{fulbright-mcwilliam-rich+07}, \citet{zoccali+06}, and \citet{lecureur+07} all came to the conclusion that the [$\alpha$/Fe] ratio was enhanced by nearly +0.1 dex with respect to the trends of both the thin and the thick disc, thus implying a shorter formation time scale for the bulge than from both discs. However, as first pointed out by \citet{melendez+08}, the bulge $\alpha$-element over-enhancement with respect to the thick disc was a result of systematic offsets between abundance measurements in dwarf stars of the disc and giant stars from the Bulge. Later on, \citet{alves-brito+10} and \citet{gonzalez+11} confirmed that when giants from the both bulge and the disc are homogeneously analysed the Bulge followed the same over-abundance in $\alpha$-elements as the thick disc, both being enhanced with respect to the thin disc at metallicities [Fe/H]$<$-0.2. At solar metallicities, the Bulge stars are found to be $\alpha$-poor, as poor as those of the thin disc. The way these trends are interpreted is that the metal-poor population of the bulge underwent a similarly fast formation scenario to that of the thick disc, while the metal-rich population of the bulge must have had a longer formation time scale, in similar time-scale to that of the thin disc stars. Similar conclusions have been reached in several other studies carried in different regions of the Bulge \citep{bensby+10, bensby+11, ryde+10, hill+11, johnson+11, johnson+13, johnson+14}. 

However, open questions remain regarding the chemical similarities of Bulge stars with those of the thick disc stars, particularly in light of a few recent findings. \citet{bensby+13} suggested that the position in the [$\alpha$/Fe] -- [Fe/H] plot where [$\alpha$/Fe] starts to decrease (referred to as the \textit{knee} in the literature) is located at higher metallicities in the Bulge than in the thick disc. The position of the \textit{knee} in the bulge may be 0.1--0.2 dex higher in metallicity in the Bulge than in the thick disc thus suggesting that the chemical enrichment of the metal-poor bulge has been somewhat faster than what is observed for the local thick disk. As the sample of Bulge micro-lensed dwarfs increases, it would be of great interest to further confirm the findings of \citet{bensby+13}. As a matter of fact, a similar result was proposed by \citet{johnson+14} who also added the analysis of Fe-peak elements finding in particular that Co, Ni, and Cu appear enhanced compared to the disc. It is important to recall that the results presented in \citet{johnson+14} have been obtained by comparing Bulge giants to dwarf stars from the local disc.  This technique has been shown to suffer from systematic offsets \citet{melendez+08}. However, the detailed analysis by \citet{johnson+14} has been carefully calibrated internally so it would be of great interest to confirm if these results are also found when bulge giant stars are compared to (inner) disk giant stars. These findings certainly highlight the importance of the future multi-object spectroscopic surveys on different galactic components to obtain a definitive answer.  

\section{The Bulge Kinematics}

The observational properties of the Bulge regarding the morphology, age and chemical abundances seem to be independently providing evidence for a rather complex bulge stellar population, where at least a metal-poor, alpha-enhanced population of old stars co-exists with a metal-rich, alpha-poor population of both old and a fraction of young stars ($\sim22$\% according to \citet{bensby+13} but see \citet{clarkson+11}). It is then natural to evaluate how the kinematics of each of these populations can help to solve the puzzle and answer the question on what is the nature of each of these components.

Following the initial attempts to constrain the kinematics of the bulge stars by measuring their radial velocities \citep{frogel+87, rich+88, rich+90, terndrup+95, minniti+96, sadler+96, tiede-terndrup+97} our understanding of the dynamical characteristics of the Bulge has gained an outstanding level of advancement thanks to the recent spectroscopic surveys that are able to sample radial velocities of thousands of M giants and red-clump stars across the Bulge.

In the most general view of Galactic bulge kinematics, the bulge is known to lie between a purely rotating system and a hotter system supported by velocity dispersion, with a $V_{max}$/$\sigma = 0.65$ \citep{minniti-zoccali+08}. The Bulge Radial Velocity Assay survey \citep[BRAVA;][]{howard+08,kunder+12} was the first to provide a broad view of the bulge kinematics, perhaps finally allowing us to start looking at the Bulge from an extragalactic perspective. The BRAVA survey presented the mean radial velocity and velocity dispersion as a function of longitude, at different Bulge latitudes, showing evidence for cylindrical rotation of the bulge \citep[BRAVA; ][]{howard+08,kunder+12} which is a characteristic feature of box/peanuts originating from secularly evolved bars. 

%
\begin{figure}[]
\sidecaption
\includegraphics[angle=0, scale=0.66,trim=0cm 0.0cm 0.0cm 0cm,clip=true]{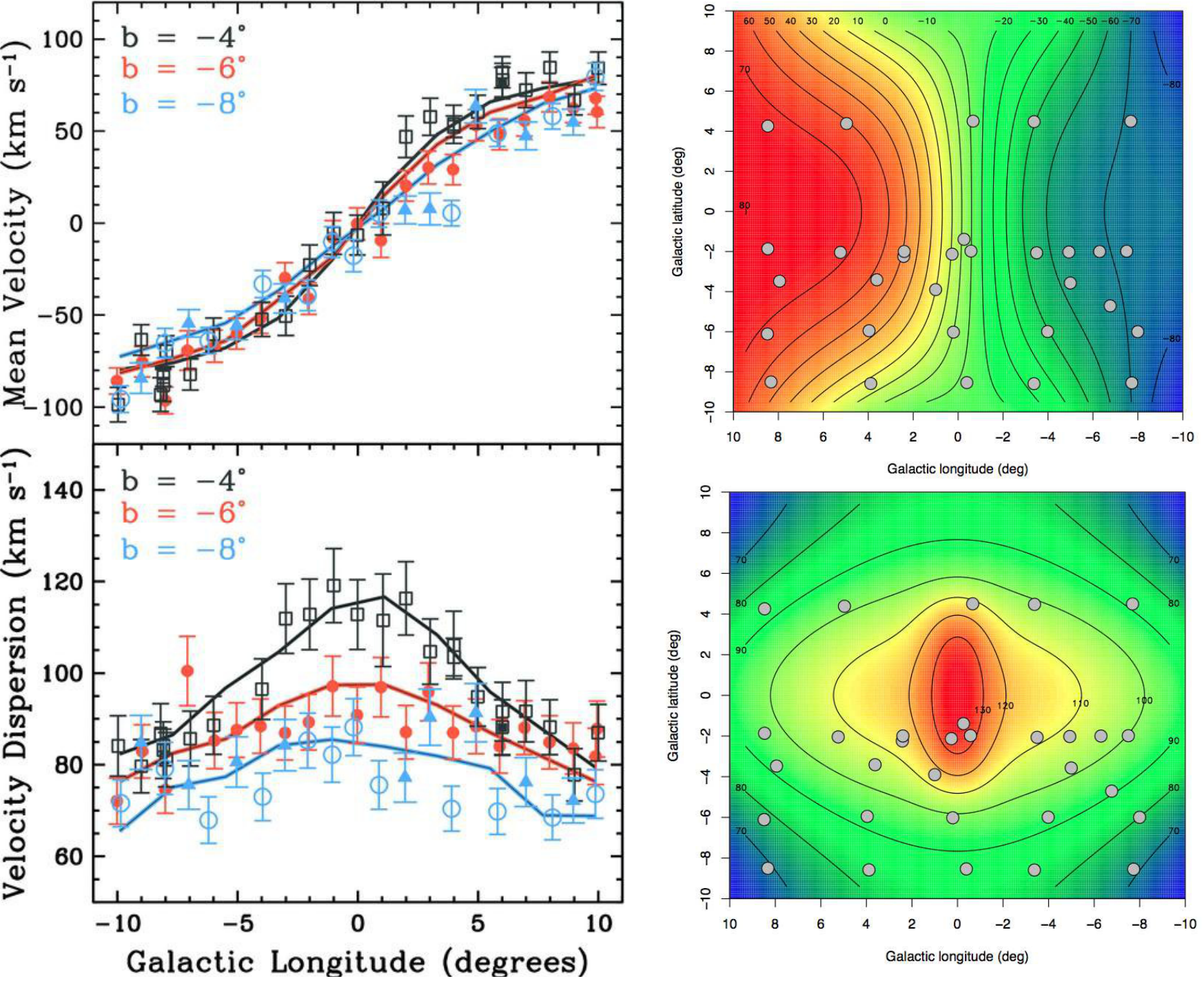}
%
%
\caption{Left panels: Velocity dispersion profile (bottom) and rotation curve (top) for latitudes b = -4, -6, and -8 from \citet{kunder+12}. Solid lines represent the models from \citet{shen+10}. Right panels: Radial velocity (top) and radial velocity dispersion (bottom) surface in the longitude-latitude plane constructed from the measured rotation profiles at negative latitudes from the GIBS survey by \citet{zoccali+14}. Grey points show the positions of the observed fields by the survey, while the black contour lines are labeled with the relevant velocity dispersion in km/s.  [{\it Left panel adapted from Fig. 11 in \citet{kunder+12}, \copyright AAS reproduced with permission. Right panels adapted from Fig. 10 and 11 in \citet{zoccali+14}, \copyright ESO reproduced with permission.}]}
\label{rv_figs}       
\end{figure}

The cylindrical rotation seen in M giants of BRAVA, later confirmed using red-clump stars by the ARGOS \citep{ness-kine+13} and GIBS \citep{zoccali+14} surveys, was modelled by \citet{shen+10}, who conclude that the bulge could not have any more than $8\%$ of the disc mass in the form of a classical spheroid to reproduce BRAVA observations. Figure~\ref{rv_figs} shows the agreement between the BRAVA survey measurements at different latitudes compared to the pure disc models from \citet{shen+10}. Also shown in Fig.~\ref{rv_figs} are the radial velocity and velocity dispersion maps for the Bulge constructed by \citet{zoccali+14} based on the GIBS survey observations.

\begin{figure}[]
\sidecaption
\includegraphics[angle=0, scale=.33,trim=2.5cm 0.0cm 0.0cm 0cm,clip=true]{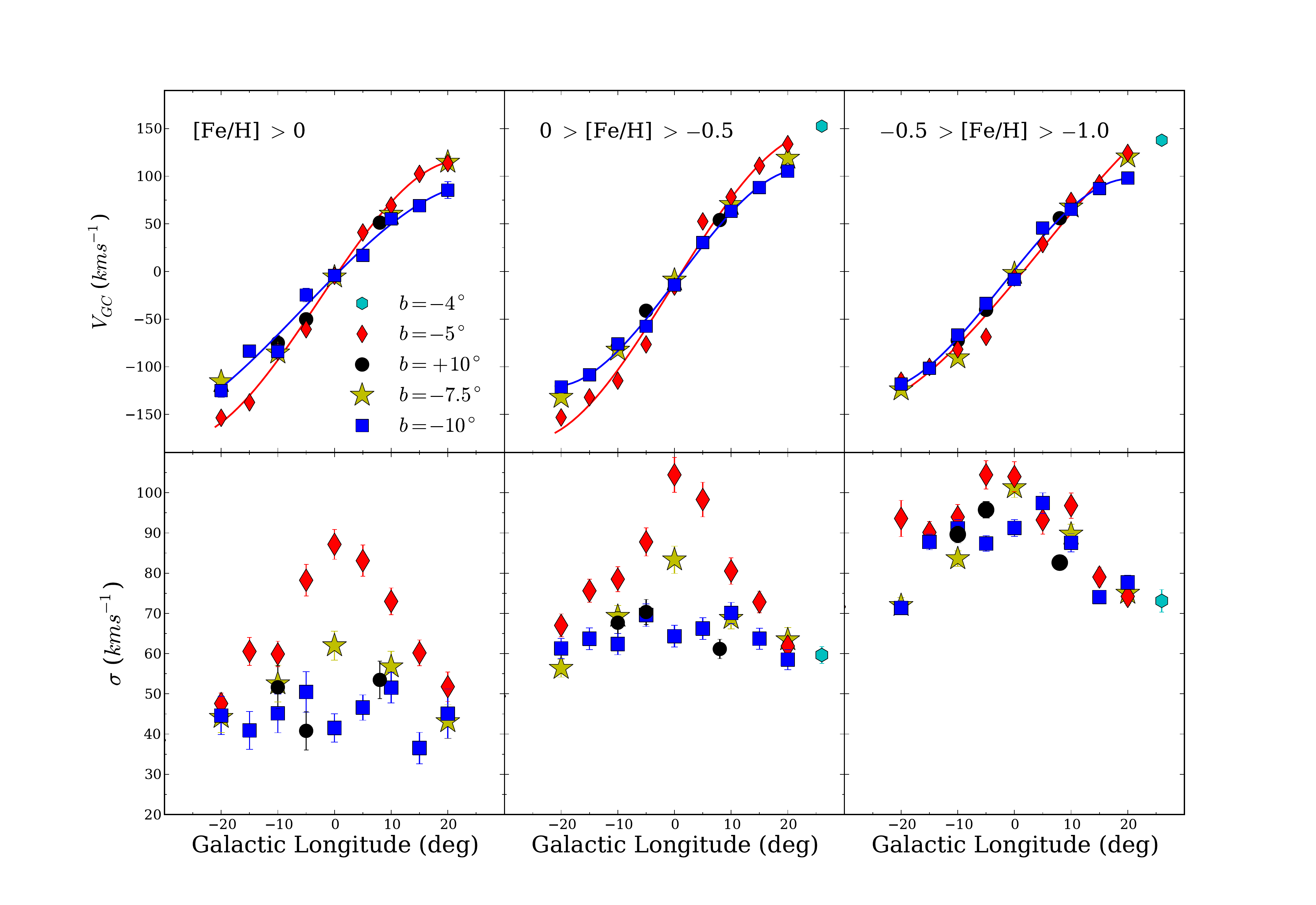}
%
%
\caption{Rotation (top panel) and velocity dispersion (bottom panel) across the bulge for the 16600 stars from the ARGO survey with [Fe/H] $>$ -1.0 within $<$ 3.5 kpc of the Galactic Centre from \citet{ness-kine+13}. The three plots correspond to different metallicity bins, from left to right in decreasing [Fe/H]. Note that the discrete bins are used to represent stars of components A, B and C from left to right shown in Fig.\ref{argos_mdf}. Although the rotation curves are similar, the dispersion clearly demonstrates the difference in kinematics of stars with [Fe/H]$>$ -0.5 and with [Fe/H]$<$ -0.5. There are 3100, 8600 and 4900 stars in each plot, from left to right. The red diamonds are b = -5, the yellow stars are b = -7.5, the blue rectangles are b = -10 and the black circles are b = +10.  [{\it Figure adapted from Fig. 6 in \citet{ness-kine+13}.}]}
\label{rv_argos}       
\end{figure}

However, dynamical models such as the one presented by \citet{saha+12} have shown that if a spheroidal component, i.e. a classical bulge, was already present when the bar was formed then the classical bulge could spin-up and rotate faster than expected for its dispersion supported nature due to the effects of the bar potential \citep{saha+13}. In these conditions, the detection of such a component based in kinematics alone would be very difficult \citep{gardner+14}. Indeed, in order to understand the nature of these components the analysis of a connection between the kinematics and other stellar properties such as metallicity seems to be a key factor. \citet{babusiaux+10} investigated the connections of metallicity and kinematics, the latter based on radial velocities and proper-motions, for the sample of \citet{zoccali+08} at different latitudes along the bulge minor axis. They found that the high metallicity stars ([Fe/H]$>$-0.25) show a larger vertex deviations of the velocity ellipsoid than their metal-poor ([Fe/H]$<$-0.25) counterpart. Furthermore, metal-rich stars showed an increase in their velocity dispersion with decreasing latitude (moving closer to the galactic plane), while metal-poor stars show no changes in the velocity dispersion profiles. This information led \citet{babusiaux+10} to associate the more metal-rich stars with a barred population and the metal-poor stars with a spheroidal component or even the inner thick disc. The rotation curves and dispersion profiles of the large sample of stars from the ARGUS survey (Fig.~\ref{rv_argos}) led \citet{ness-kine+13} to reach a similar conclusion to that of \citet{babusiaux+10}. 

%

%
\begin{figure}[]
\sidecaption
\centering
\includegraphics[angle=0, scale=1.2,trim=0cm 0.0cm 0cm 0cm,clip=true]{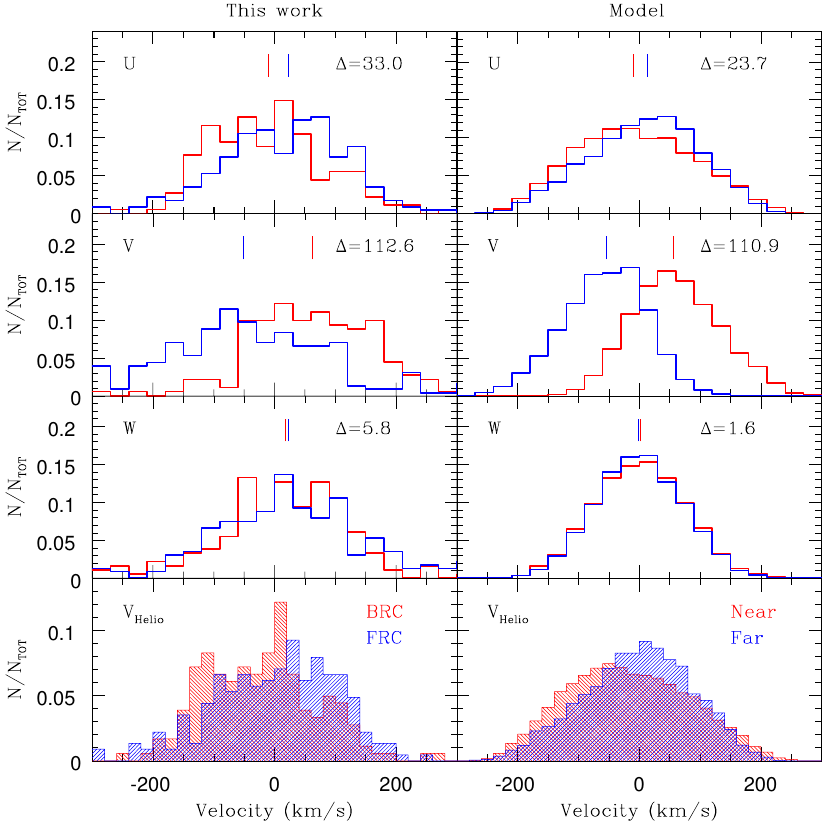}
%
%
\caption{Comparison of the kinematics of the bright and faint red clump of the Bulge with respect to the kinematical model for a strong boxy-peanut bulge \citep{debattista+05} obtained from \citet{vasquez+13}. From the model, two samples were selected from the two overdensities formed by the near (red) and far (blue) arms of the boxy-peanut stellar distribution in the line of sight for (l,b) = (0, -6). Colour lines over U, V, and W histograms correspond to the median value for each distribution. [{\it Figure adapted from Fig. 7 in \citet{vasquez+13}, \copyright ESO reproduced with permission.}]}
\label{ppm_fig}       
\end{figure}

An additional piece of the puzzle has been provided thanks to the recent development of our understanding of the structural properties of the Bulge, in particular the discovery of the X-shape. \citet{ness-kine+13} and \citet{vasquez+13} investigated in more detail the connection between the X-shape bulge and its chemo-dynamical properties. Both studies showed that only the metal-rich stars ([Fe/H]$> -0.5$) trace the split red-clump in the luminosity function and therefore belong to the X-shaped bulge, while the metal-poor stars do not share the same split in magnitude.

Finally, important constraints on the bulge formation history can be obtained by looking at the radial velocities of a large number of stars. However, also having information on the proper motions for some of those stars in order to reconstruct the three-dimensional characteristics of the stellar orbits may become fundamental to disentangle the more complex characteristics of the Bulge. Although proper motions can be derived for large samples of stars based on photometric information obtained with relatively short exposure times, they require a long time baseline ($\sim10$ yr). Therefore, to obtain a very precise astrometric solution in order to reach the required accuracy of a few mas/yr across the bulge is perhaps more a luxury than a requirement. Indeed, studies that have used proper motion information in specific fields, combined with radial velocity measurements, have been able to unravel the perhaps otherwise hidden complexity of the Bulge \citep[e.g.][]{zhao+94, soto+07, babusiaux+10}. In this context, \citet{vasquez+13} provided for the first time an analysis based on both the radial velocities and proper motions for both arms of the X-shape bulge, and were thus able to derive the complete space velocities in the U, V, W Galactic Cartesian system for a sample of spectroscopic targets in the field (l,b)=(0,-6). As shown in Fig.~\ref{ppm_fig}, \citet{vasquez+13} showed how the closer over-density of the X-shaped bulge shows an excess of stars moving towards the Sun, and the far overdensity shows an excess of stars receding from the Sun, as expected from stars on elongated orbits streaming along the arms of the X-shaped bulge. A wider mapping of these are the key signatures of the detailed kinematics of the bulge, thus expanding the study of \citet{vasquez+13} to other Bulge regions, will become an important ingredient for an accurate modelling of the bulge and will perhaps allow a better identification of its different components \citet{gardner+14}.

\section{The Milky Way bulge in a nutshell}

The properties of the Galactic bulge described above can be summarised as follows:

\begin{itemize}

\item The Bulge stellar population properties show independent evidence for a multiple component scenario, with different morphological and dynamical characteristics. The metallicity distribution can be separated in, at least, a main metal-poor and a metal-rich components. Their different contributions across the Bulge produces a vertical metallicity gradient of $\sim$0.6 dex/kpc which seems to flatten in the inner 700 pc. From a morphological/structural viewpoint, there is strong evidence of a box/peanut, while more recent work points out the existence of a component with different geometrical properties, as seen in the spatial distribution of RR Lyrae stars, which trace the oldest Galactic population (Fig. \ref{rrlyrae_dist}). The next two bullet points connect these structural components to differences observed also in chemical content, age and dynamics.

\item There is a metal-rich (mean [Fe/H]$\sim$0.3), alpha-poor population of stars in the bulge, composed mostly by old stars but with a fraction of young stars. This population of stars rotates cylindrically and shows a large vertex deviation consistent with the bar structure traced with a position angle of $\sim$27 deg. At latitudes $|b|>5$ the inner parts of the bar have grown out of the disc plane, originating the box/peanut. These metal-rich stars further show the split red-clump in the luminosity function tracing the X-shape of the Bulge.

\item The metal-poor bulge population (mean [Fe/H]$\sim$-0.3) is composed predominantly by old stars and it shows an alpha-enhancement similar to that of the local thick disc. The kinematics of these stars follow a more spheroidally distributed population than the one traced by the metal-rich stars, consistent with the structure traced by the bulge RR Lyrae stars. Furthermore, these stars do not trace the X-shape morphology of the Bulge.

\end{itemize}

\section{The Milky Way bulge in the context of external galaxies}

\subsection{The X-shaped Bulge of the Milky Way: how rare is this?}

It can be argued that while in the context of Galactic research box/peanuts are somewhat a recent discussion, in Extragalactic studies such structures are known for about twice as much the time. So it came as no surprise to the Extragalactic community when evidence suggested that the Milky Way has a box/peanut, especially because there is also evidence that it has a bar. Perhaps the first mention in the literature about these deceptively unusually-looking structures is from \citet{BurBur59}, referring to the prototypical example that is NGC 128 \citep[see also][]{San61}. More detailed investigation came with \citet{deV74}, \citet{Jar86} and \citet{Sha87}. But it was in the pioneering studies of \citet{deSDos87} that the major step of connecting these structures with bars was made for the first time from an observational viewpoint, using a statistical argument. Basically, they argued, the frequency of box/peanuts in edge-on lenticulars is similar to that of bars in face-on lenticulars, consistent with the idea that box/peanuts are bars seen at a different projection. This conclusion was corroborated years later by \citet{LueDetPoh00}, who reported a fraction $>40$ per cent of box/peanuts in disc galaxies covering most of the Hubble sequence (from S0 to Sd classes).

The complicating factor here is of course the fact that bars are difficult to be seen when the inclination angle of the galaxy is too large. Therefore, simulations of barred galaxies played a major role here. In fact, the starting point for this observational connection between bars and box/peanuts was the work published in \citet{ComSan81}. These authors have shown, using collisionless simulations, that bars seen at a given edge-on projection show a very characteristic peanut-like morphology.

A number of studies came thereafter dedicated to extend this connection into a dynamical context. \citet{KuiMer95} came up with an ingenious diagnostic to test in this context whether box/peanuts are just bars seen at a different projection. This consisted in producing diagrams in which the line of sight velocity is plotted against the galactocentric radius for highly inclined or edge-on systems. By producing such diagrams corresponding to orbits in a purely axisymmetric potential, and orbits in a barred potential, they showed that the presence of a bar produces a clear distinctive signature. Because at a region around the bar corotation radius (the radius at which the pattern speed of the bar matches the local circular speed) there are no close, non-self-intersecting orbits available, clear gaps appear in this diagram, producing a figure-of-eight pattern. The matter became then just to produce such diagrams for galaxies presenting box/peanuts in order to test for the presence of a bar. \citet{KuiMer95} did that for NGC 5746 and NGC 5965, providing observational evidence that box/peanuts and bars are related phenomena.

More evidence was produced in \citet{MerKui99} and \citet{BurFre99}. To this point, almost all galaxies with box/peanuts studied showed evidence of a bar. In only a few extreme cases the box/peanut could have formed through accretion of external material. In addition, none of the galaxies without box/peanuts showed signatures of a bar. Further development also happened in the theoretical background. \citet{BurAth99} refined and corroborated the orbital study of \citet{KuiMer95}, while \citet{AthBur99} provided strong support to the bar detection diagnostic with hydrodynamical simulations.

It must be noted that the detailed morphology of box/peanuts, i.e. if they either have a boxy shape, a peanut shape or an X shape, depends on projection effects, as well as the strength of the box/peanut. More about X shape bulges can be found in Laurikainen \& Salo (this volume), but essentially the X shape is more clear in the strongest peanuts. Techniques such as unsharp masking are able to reveal X shapes more clearly when the peanut is not so pronounced. Therefore, all these morphologies come actually from the same physical process, i.e. the increase in the vertical extent of stellar orbits in the inner parts of the bar. Both simulations and observations point out that box/peanuts extend to galactocentric distances which are about a third to a half of the bar semi-major axis \citep[see][]{ErwDeb13}. What causes this change in the orbits vertical extent is reviewed in detail by Athanassoula (this volume). In the Milky Way, we see the extension of the B/P to galactocentric distances of $\sim1.5$ kpc, which is nearly two-thirds the length of the semi-major axis of the long bar of 2-2.3 kpc \citep{hammersley+00, benjamin+05, lopez-corredoira+07}.

\begin{figure}
\begin{center}
\includegraphics[scale=0.15,trim=0cm 2cm 0cm 0cm,clip=true]{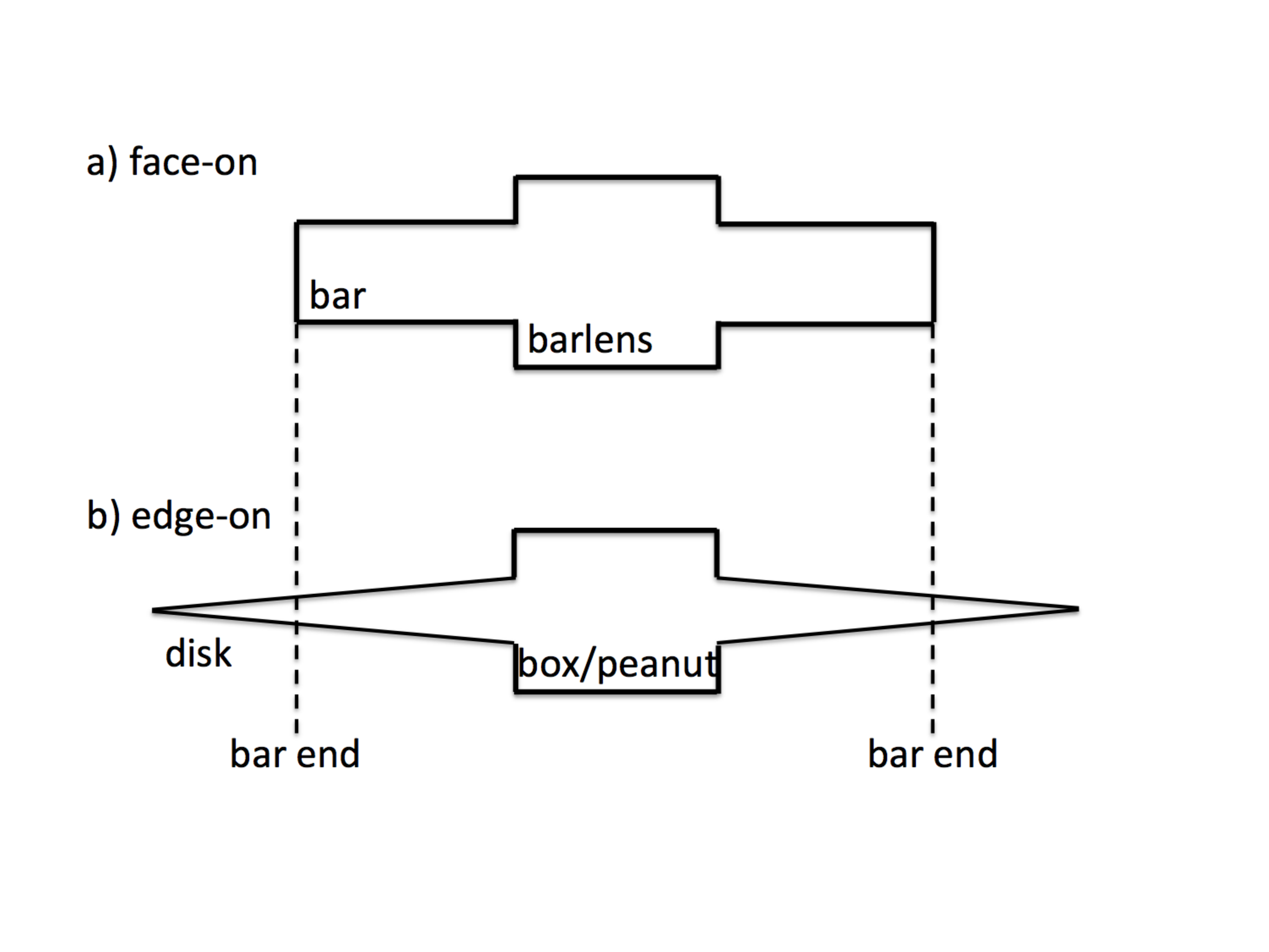}\includegraphics[scale=0.25,trim=13cm 0cm 0cm 0cm,clip=true]{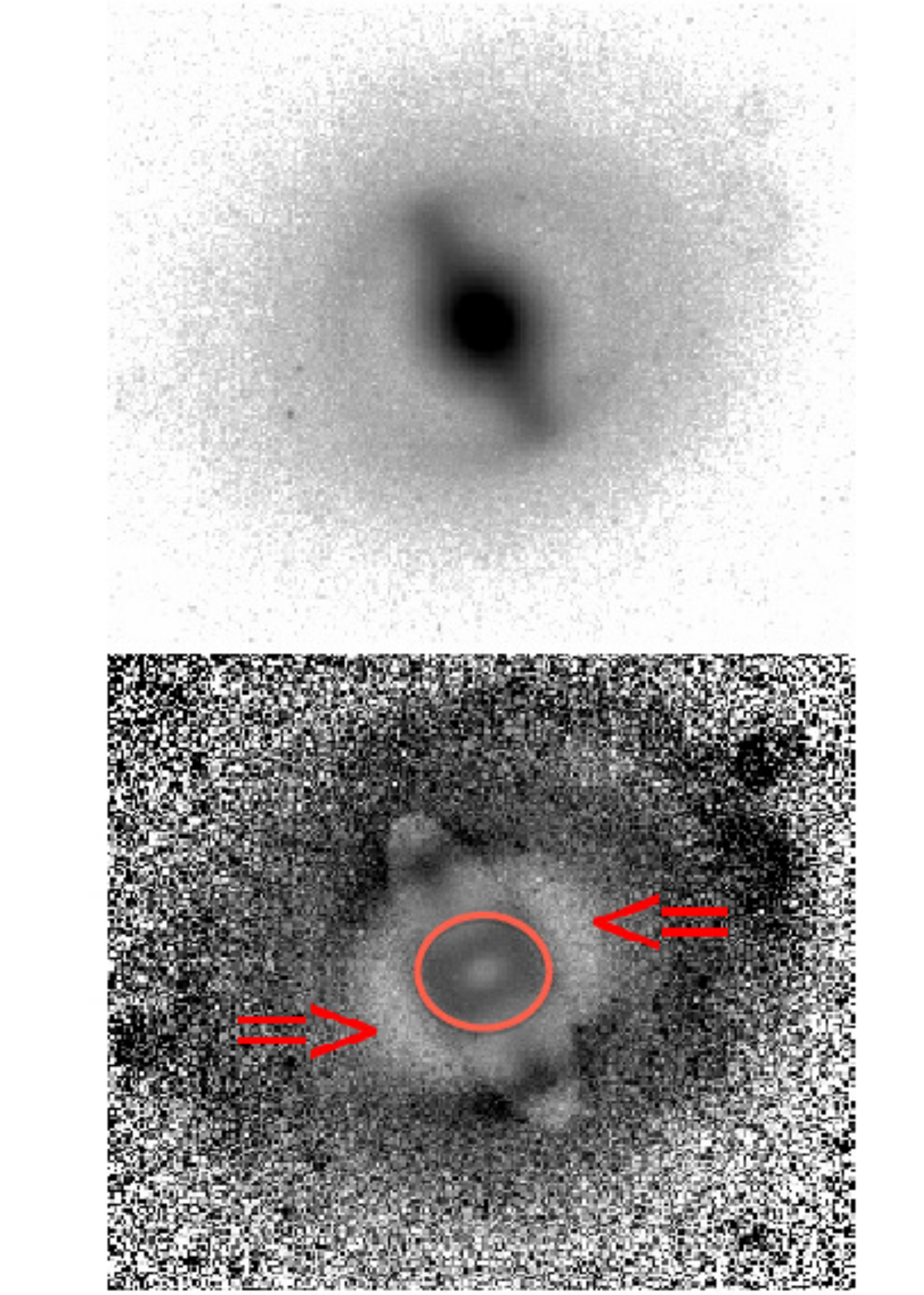}
\end{center}
\caption{Connection between bars, box/peanuts and barlenses. The diagram on the left shows a schematic representation of a barred galaxy seen a) face-on, and b) edge-on. Box/peanuts can be seen at edge-on projections, but the flat, more extended part of bar is difficult to realize, as its vertical extent is similar to that of the disc. The same galaxy seen face-on would reveal the bar and barlens. The barlens and the box/peanut appear to be the same structure seen at different projections [see Laurikainen \& Salo (this volume) and Athanassoula (this volume)]. The panels on the right show an $R$-band image of NGC 4608 (top), where the bar and barlens can be clearly seen \citep[see also][]{LauSalBut05,LauSalBut07,LauSalBut11}. The bottom panel shows a residual image, after the subtraction of a 2D model of the bulge, bar and disc of this galaxy. In the residual image the barlens stands out even more clearly. The red circle points out the barlens. This circle was not in the
original Gadotti (2008) paper, but added now that we understand that
the structure is a barlens. The red arrows point out empty regions in
the disc within the bar radius, where stars from the disc were
captured by the bar. [{\it Right panels adapted from \citet{Gad08}.}]}
\label{fig:barlens}
\end{figure}

A review on the observed properties of box/peanuts in external galaxies is given by Laurikainen \& Salo (this volume). They also discuss a structure called barlens, which is interpreted as the projection of box/peanuts when seen face-on. This structure was noticed by \citet{LauSalBut05}, who included a model to fit barlenses in their image decompositions. Later, based on Fourier analysis, \citet{LauSalBut07} suggested that barlenses are part of the bar, while \citet{Gad08} also noticed their existence in a sample of local barred galaxies. However, only in \citet{LauSalBut11} the term 'barlens' is introduced as a new morphological feature in galaxies. Very recently, more detailed studies have made a robust connection between barlenses and box/peanuts \citep{LauSalAth14,AthLauSal14}. This connection implies that also on the plane of the disc, the stellar orbits in the inner part of the bar become wider. Figure \ref{fig:barlens} describes schematically the connection between bars, box/peanuts and barlenses.

The presence of barlenses in external galaxies suggests of course that our own Milky Way may have such structural component. Since the stars in barlenses seem to be contained within the disc plane, in the Milky Way, they are seen in projection, in the foreground and background with respect to the Galactic centre. They thus complicate even further the interpretation of observations of the Milky Way, such as those discussed above. These stars are stars within the bar and then have chemical properties and ages similar to those of other bar stars -- they are box/peanut stars. From a kinematical point of view, barlenses are different from discs and this is a promising avenue to separate them from the other stellar populations seen from the Sun at the direction of the Milky Way central regions.

\subsection{Bulge formation scenarios}

The formation of bulges in general, and of the Milky Way bulge in particular, have been discussed many times elsewhere \citep[e.g.][see also Bournaud, and Brooks \& Christensen, this volume]{AguBalPel01,KorKen04,Ath05b,LauSalBut07,HopBunCro10,FisDro10,Gad12}. Here we will assess each bulge formation scenario in the light of evidence obtained from data on the Milky Way bulge, as presented above. There is mounting evidence that the Milky Way has a bar and a box/peanut, and thus secular evolution processes induced by bars in disc galaxies must have played a non-negligible role in the evolution of the Galaxy. On the other hand, a scenario in which the Galactic bulge was formed in violent processes such as mergers has weak support from data. In fact, an important question that observers must focus now is whether the Galaxy has a merger-built central stellar component at all.

\subsubsection{Bulges formed via disc instabilities}

Dynamical disc instabilities can originate bars, spiral arms and ovals, the latter being just a distortion in the disc stellar orbits that make them acquired less circular orbits, but not as eccentric as those in bars. All these structures, being non-axisymmetric, produce perturbations in the galaxy potential, with the result that material (gas and stars) within the corotation resonance radius loses angular momentum, whereas material outside the corotation radius absorb this angular momentum. The effect is particularly important for the collisional gas component, which thus falls towards the centre. At some point, the in-falling gas gets compressed and form stars, contributing to a rejuvenation of the stellar population in the central regions \citep[e.g.][]{GadDos01,Fis06,CoeGad11,EllNaiPat2011}.

\begin{figure}
\begin{center}
\includegraphics[scale=0.33,clip=true]{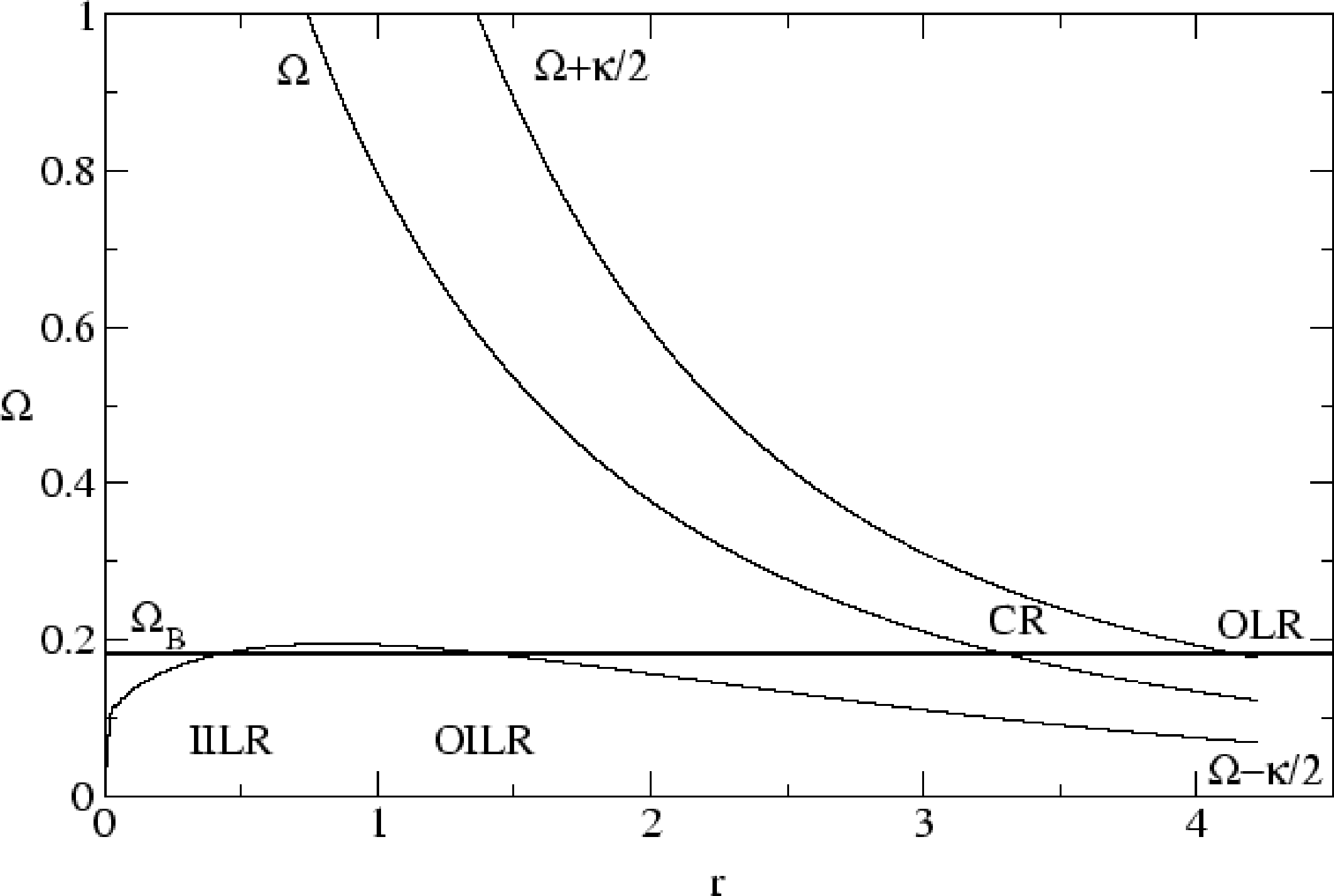}
\end{center}
\caption{Angular speed of stars in circular orbits in a potential reproducing a disc galaxy with a bar, as a function of the galactocentric distance (in arbitrary units). The bar pattern speed is represented by the solid horizontal line ($\Omega_B$). The epicyclic frequency of the stellar orbits is denoted by $\kappa$. Barred galaxies present several dynamical resonances. This figure shows schematically how four of the main resonances come to be. Whenever the bar pattern speed is equal to $\Omega$ or $\Omega\pm\kappa/2$ this is the position of a dynamical resonance. From the centre outwards, the resonances depicted here are: the Inner Inner Lindblad Resonance, the Outer Inner Lindblad Resonance, Corotation, and the Outer Lindblad Resonance. In this case, the main families of stellar orbits change their orientation by 90 degrees at each resonance. This effect is at the origin of a number of dynamical effects in barred galaxies, in particular the transfer of angular momentum from material inside the corotation radius to material outside this radius, and the resulting formation of disc-like bulges.}
\label{fig:resonances}
\end{figure}

Because most disc galaxies with bars should have one or two Inner Lindblad Resonances (ILRs) near the centre ($\sim$ a hundred to a few hundred parsecs from it), the in-falling gas cannot reach the galaxy centre immediately (see Fig. \ref{fig:resonances}). Instead, it usually forms an inner disc, decoupled from the large-scale disc. These inner discs may form nuclear bars and spiral arms, as often observed, and are often called disc-like bulges \citep[see e.g.][]{Ath05b,Gad12} or discy pseudobulges, to contrast with the fact that box/peanuts are often called as well pseudobulges \citep[see][]{KorKen04}. At the ILRs gas can get accumulated and compressed, often forming a star-bursting nuclear ring.

However, it must be noted that simulations indicate that, initially, the infall of gas within a bar corotation radius occurs rapidly after the formation of the bar, with a time scale of the order of 10$^8$ years, i.e. a dynamical time \citep{Ath92b,EmsRenBou14}.\footnote{Gas outside corotation receives angular momentum from the bar, and other factors govern the gas infall rate at these outer radii, such as dynamical effects induced by spiral arms and the dissipative nature of the ISM. At these distances the infall of gas no longer occurs in a dynamical time scale.} This means that disc-like bulges not necessarily present ongoing star formation or a very young stellar population, if the bar has formed long ago and have been able to push most of the gas to the centre quickly, and if the gas content in the disc is not being replenished. \citet{SheElmElm08,SheMelElm12} present results that suggest that the first long-standing bars\footnote{Some simulations \citep{KraBouMar12} have reported the early formation of bars, at redshifts above 1. However, these bars are short-lived. In these simulations, bars formed at $\leq 1$ generally persist down to $z = 0$.} formed after redshift 1. These bars are still the ones seen at redshift zero, since bars are difficult to destroy, unless the disc is extremely gas-rich \citep[see][]{AthLamDeh05,BouCom02,BouComSem05,KraBouMar12}. This means that they first induced star formation at the centres of their host galaxies about 8 Gyr ago. Thus, disc-like bulges with stars as old as 8 Gyr are perfectly possible.

For the disc to be replenished with gas so that the bar can push this gas to the centre and produce a new central burst of star formation and rebuilding of the disc-like bulge, it has to fall into the disc plane from a direction not parallel to the galaxy disc, and inside the bar corotation radius. Otherwise, this gas will be pushed outwards by the bar or accumulate at the corotation \citep[see][]{BouCom02}. Evidence for gas infall from directions not parallel to the galaxy disc has recently been presented by \citet{BouMurKac13}, but how often it occurs is still unknown, as is whether the gas reaches the disc within corotation.

\begin{figure}
\begin{center}
\includegraphics[scale=0.27,clip=true]{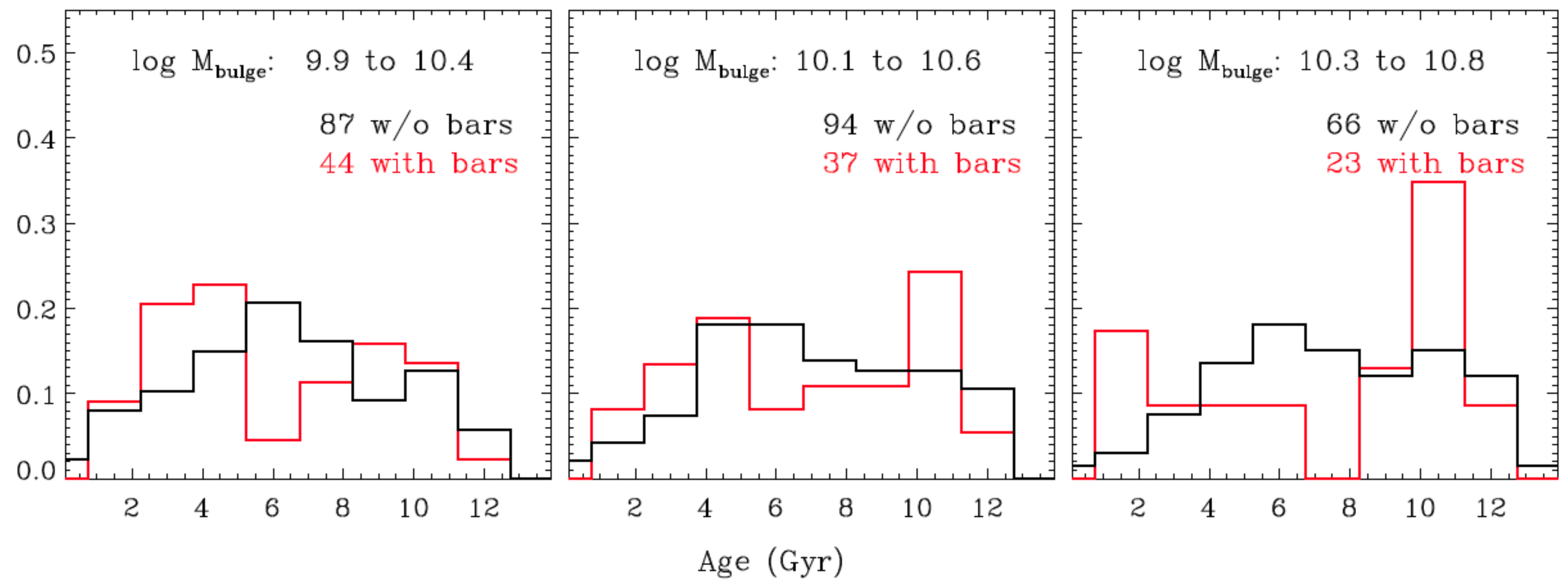}
\end{center}
\caption{Relative distributions of bulge mean stellar ages for barred and unbarred galaxies in bins of bulge stellar mass, as indicated. For massive bulges, the distribution is bimodal only for barred galaxies, consistent with the picture in which secular evolution processes build disc-like bulges. However, note that the mean stellar ages of such bulges can be as high as a few Gyr. [{\it Adapted from \citealt{CoeGad11}.}]}
\label{fig:CoeGad11}
\end{figure}

Thus, although \citet{EllNaiPat2011} find that, statistically, barred galaxies present ongoing, central star formation more often than unbarred galaxies, there is still a significant fraction of barred galaxies with star formation rates comparable to those in unbarred galaxies. In addition, \citet{CoeGad11} find that the younger bulges found in barred galaxies have a mean stellar age of a few Gyr. This is in contrast to unbarred galaxies, which show on average older mean stellar ages (see Fig. \ref{fig:CoeGad11}). This means that replenishing the disc with gas inside the corotation radius is a phenomenon that does not occur very often. Otherwise, very young stellar populations should be more conspicuous at the centres of barred galaxies. As discussed above, \citet{BenYeeFel13} find a stellar component in the Milky Way bulge with ages less than 5 Gyr. Although this means that the mean stellar age for the Milky Way bulge as a whole is above this value, this younger component has a mean stellar age thus in very good agreement with the mean stellar ages of the young bulges in other barred galaxies. The bottom line is that stars originating from gas infall to the centre through disc instabilities do not necessarily have to be extremely young now. A fraction of these younger stars could be elevated out of the disc plane and populate the box/peanut, but most of these stars are expected to be at or near the large scale disc plane.

Some studies using both simulations and observations have suggested that some star formation may occur along the bar as soon as the bar forms and induces shocks in the gas content in the interface between the bar and the disc \citep[see][]{Ath92b,Phi96,SheVogReg02}. However, they also indicate that, due to shearing in the gas clouds when they start falling towards the centre, this is interrupted shortly afterwards, and from then on, star formation is limited to components at the bar ends and the disc centre. Discs stars trapped by the bar mostly do not leave the bar, keeping their elongated orbits. As it evolves, however, the bar can capture stars from the disc formed more recently (see an example of such capture in the right panels of Fig. \ref{fig:barlens}), and thus a fraction of young stars can be present in the bar, as long as there is ongoing star formation in the disc within the radius at which the bar ends, and the bar can grow stronger and keep capturing disc stars. Stars in bars are thus predominantly old, and therefore box/peanuts, being just part of bars, should as well be populated mostly by old stars, with some younger component.

For a galaxy as massive as our own, the bar is expected to form at redshifts close to 1 \citep[see][]{SheMelElm12}. The ages of the stellar populations seen at the Galactic bulge are thus consistent with the picture of it being built purely from the bar instability in the disc, i.e. the Galactic bulge can well be just a box/peanut plus a disc-like bulge, as far as the ages of its stellar populations is concerned.

In trying to assess how the bulge of the Galaxy has formed, the chemical content of its stellar populations are better consider closely with their kinematical properties. Bars and their box/peanuts are expected to rotate cylindrically, i.e. the mean stellar velocity is independent of the height above the plane of the disc -- as a rigid body. Variations in this pattern are usually attributed to the presence of other structural components with different kinematical properties \citep[see e.g.][and references therein]{WilZamBur11}. As discussed above, one stellar population in the Bulge can be described as having high metallicity, low content of alpha-elements, and kinematics consistent with the eccentric stellar orbits in bars in cylindrical rotation. The alpha-element content thus indicates that the box/peanut formed after the thick disc. In external galaxies, recent evidence suggests that thick discs form early, during the short initial formation stages of the galaxy, which qualitatively agrees with the picture for the Milky Way \citep{ComElmKna11,ComElmSal14}. Nevertheless, although most of the thick disc stars thus form in situ, a significant fraction of stars in the thick disc may come from the accretion of satellites, and another fraction (likely smaller), may consist of heated up stars from the thin disc. Evidently, this complicates substantially the interpretation of observations.

Likewise, the observations from the ARGUS survey indicating variations in metallicity within the box/peanut (see Sect. 4.1), are suggestive of a complex stellar population content in the box/peanut itself. This can be a result of similar processes that complicate just as well the stellar population content of the thick disc, as described in the previous paragraph. As discussed in Sect. 4, different populations or population gradients in the disc from which the box/peanut forms can produce a similar result. However, it can also be the result of more than one buckling event forming the box/peanut. In the simulations of \citet{MarShlHel06} a bar goes through a first buckling event about 2 Gyr after the beginning of the simulation, and this event is fast ($<1$ Gyr). However, a second, powerful buckling event occurs around 5 Gyr later and lasts for about 3 Gyr. How this would affect the chemical content of stars seen today in the box/peanut depends on how the stellar population content and kinematical properties of the bar and box/peanut vary during these periods. But it is clear that if the bar of the Milky Way has gone through such recurrent buckling, the presence of multiple populations in the box/peanut is not surprising. Add this to the complex composition of the thick disc and one sees how complicated the stellar populations can be away from the disc plane.

\subsubsection{Bulges formed via violent processes}

The classical picture of bulges in disc galaxies is that of mini-ellipticals: massive, smooth and extended spheroids with dense centres, with old stellar populations showing alpha-enhanced chemical composition and relatively low rotational support (as compared to discs). A natural formation scenario for these structural components would be that of a monolithic collapse \citep{EggLynSan62}, in which a single gas cloud collapses in short time scales ($< 10^8$ yr), producing a violent burst of star formation that originates the stellar halo and the bulge. While this scenario might explain the formation of the first spheroids, it faces many difficulties. It does not reproduce for instance the heterogenous distributions of stellar ages and metallicities observed in bulges \citep[e.g.][a monolithic collapse implies a more homogeneous population]{McWRic94,WysGilFra97} and regions of ongoing star formation \citep[e.g.][]{CarStideZ97}.

It should be noted, however, that the monolithic collapse scenario was formulated within a perspective that does not include box/peanuts, and thus should not be compared against the properties of such observed structures. On the other hand, modern dissipative collapse models similar to the pioneer model of \citet{EggLynSan62}, that however include as well cosmological ingredients, generate bulges with more realistic properties. Such key ingredients include in particular a long time scale history of accretion of dark matter haloes into the central halo, with the associated evolution of angular momentum and star formation episodes \citep[see e.g.][]{SamGer03,ObrDomBro13}.

A natural picture within $\Lambda$CDM cosmology is that of merger-built bulges. Brooks \& Christensen (this volume) review this picture. Another scenario to explain the formation of classical bulges is the coalescence of clumps in discs at high redshifts. This is also reviewed in this volume by Bournaud.

The evidence from observations of the Galactic bulge as reviewed above, however, give little support to the presence of a massive classical bulge. The observed box/peanut and its cylindrical rotation cannot be originated in violent scenarios. On the other hand, the possibility of a small classical bulge embedded within the box/peanut is not yet ruled out. Such composite bulges are discussed in the next section.

Nevertheless, we have seen above that there is a component in the Galactic bulge with low metallicity and an alpha-element content consistent with it being formed concomitantly with the thick disc, i.e. {\em before} the box/peanut. In addition, the morphological properties of this component seem to point out a more spherically distributed structure (see Fig. \ref{rrlyrae_dist}). This is revealed by RR Lyrae stars and are properties that are shared by classical bulges. This component has a spatial extent similar to that of the box/peanut, but it is not revealed by images such as those from COBE. 
However, a number of early-type disc galaxies with massive classical bulges shows bars, which probably went through a buckling process that originated a box/peanut. So it is perfectly plausible to have a classical bulge and a box/peanut coexisting in the same galaxy. A possible example is our own massive neighbor, M31. \citet{AthBea06} have shown that this galaxy has a bar and a box/peanut, and there is evidence that it also hosts a classical bulge \citep[e.g.][and references therein]{CouWidMcD11}. 2D decompositions in Gadotti \& Erwin (in preparation) show that the classical bulge has a similar extent as the box/peanut. The morphology of the M31 bulge (see Fig. 2 in \citealt{AthBea06}) is similar to that seen in the COBE/DIRBE image for the Milky Way, although the vertices of the box/peanut are more clearly recognized in the Galaxy (perhaps due to projection effects in M31). Nevertheless the same 2D decompositions of Gadotti \& Erwin reveal the X shape outstandingly \citep[see][]{Gad12}. Another critical issue is the understanding of how bright/massive is the component revealed by RR Lyrae stars in the Galaxy, and how does this compare to other classical bulges. It clearly cannot be large enough as to mask the vertices of the box/peanut revealed by COBE.

\subsubsection{Composite bulges}

In the previous subsections, we explored the possibility that the Milky Way has a disc-like bulge, apart from its box/peanut. We also remarked about the possibility of a classical bulge. Here we will briefly summarise recent work on the presence of such composite bulges in external galaxies. It is not hard to contemplate the possibility of such composite systems. A pure disc galaxy can form at high redshifts, say $z\sim3$, and acquire a classical bulge, be it through minor mergers, accretion events or the coalescence of clumps of stars and gas in the disc. At $z\sim1$ the same disc -- now hosting a classical bulge at its centre -- might become unstable to the formation of a bar, and develop one. Quickly this bar pushes gas to the central regions of the disc, originating a disc-like bulge. Give it a couple of Gyr and the box/peanut is formed. We thus end up with a galaxy containing a classical bulge, a disc-like bulge, and a box/peanut.

\citet{Gad09b} has shown evidence that 34 per cent of his disc galaxies hosting classical bulges are galaxies possessing bulges with structural properties typical of classical bulges, but with an intensity of star formation activity characteristic of disc-like bulges. While one cannot rule out the possibility that some classical bulges may present ongoing star formation, it is also plausible that many of these galaxies actually host composite bulges. Those would be composed by an extended classical bulge with an embedded disc-like bulge. Because the classical bulge dominates the disc-like bulge in terms of mass, the composite bulge shows structural properties of classical bulges. However, using the $D_n$(4000) spectral index, \citet{Gad09b} was able to realize the intense star formation in the bulge region. Such star-forming activity, according to this interpretation, occurs at the embedded disc-like bulge. Further evidence for this and other types of composite bulges is presented in \citet{MenDebCor14}.

\citet{KorBar10} report the existence of a small disc-like bulge embedded in the box/peanut of NGC 4565. \citet{NowThoErw10} argue that NGC 3368 and NGC 3489 actually have an embedded classical bulge within component(s) built via disc instabilities. Finally, very recently, \citet{ErwSagFab14} have shown further evidence of such components that appear to be embedded classical bulges.

It will thus be no surprise if the Galactic bulge is a composite bulge.

\section{Concluding remarks}

In this review we have presented a summary of current progress towards characterising the properties of the Milky Way bulge. In recent years, the spectroscopic and photometric surveys of the Bulge have provided us with the necessary tools to build a bridge connecting the detailed stellar population properties with a global view of the Galactic bulge. As a consequence, it is now becoming possible to discuss the Bulge properties as seen from an extragalactic perspective. Such a comparison, powered by the increasing number of models to which observations can now be directly compared, is the only way in which we can set the history of events that led to the properties of the Bulge we see today.

The era of surveys looking towards the Galactic bulge was born not only from our intrinsic desire to explore, but also as a response to the increasing complexity in Bulge properties revealed with previous individual observations. The need to further map its morphology, to better constrain the spatial variations of its stellar populations and subsequently connect it with their kinematics required the larger spatial coverage offered by such large-scale observations.

While the dominant B/P nature of the Milky Way bulge has now been well established, it remains to be understood if the observed stellar population properties relate solely to the same structure or if they each have a different origin. Another way to phrase this would be to ask the following question: do the metal-poor, $\alpha$-element enhanced, old bulge stars belong to a different structure than the B/P, which was formed somehow independently to the buckling instability process of the bar (i.e. as a classical bulge)? Currently, while we have important evidence from the connections between kinematics and metallicities of Bulge stars as well as their spatial distribution, we can only suspect about the presence of different components. However, creating the link between these components and the specific formation scenarios should be done with extreme caution, as a number of processes could have played a specific role at different stages of the assembly of the Galaxy. For example, the coalescence of disc clumps and the accretion of gas could have formed a thin disc, a thick disc, and even a spheroid in the centre during the early stages of formation of the Milky Way. The merging history could have also contributed to this assembly, which will depend on gas content, mass ratio and orbital parameters of the mergers, until the formation of a bar and the onset of the buckling instability took place to shape the dominant central component we see today -- the B/P Bulge. Only by carrying out an extensive comparison of all the observational properties of Bulge stars with models and external galaxies, can we constrain the importance of all these events during the formation history of the Bulge.

\begin{acknowledgement}
We are grateful to an anonymous referee for many useful comments. We warmly thank Istvan Dekany for kindly providing us the table of individual distances to the RR-lyrae that we used to produce Figure 1.
\end{acknowledgement}
\bibliographystyle{spbasic}
\bibliography{mybiblio_rev_full}

\begin{thebibliography}{159}
\providecommand{\natexlab}[1]{#1}
\providecommand{\url}[1]{{#1}}
\providecommand{\urlprefix}{URL }
\expandafter\ifx\csname urlstyle\endcsname\relax
  \providecommand{\doi}[1]{DOI~\discretionary{}{}{}#1}\else
  \providecommand{\doi}{DOI~\discretionary{}{}{}\begingroup
  \urlstyle{rm}\Url}\fi
\providecommand{\eprint}[2][]{\url{#2}}

\bibitem[{{Abadi} et~al(2003){Abadi}, {Navarro}, {Steinmetz}, and
  {Eke}}]{abadi+03}
{Abadi} MG, {Navarro} JF, {Steinmetz} M, {Eke} VR (2003) {Simulations of Galaxy
  Formation in a {$\Lambda$} Cold Dark Matter Universe. I. Dynamical and
  Photometric Properties of a Simulated Disk Galaxy}. \apj 591:499--514,
  \doi{10.1086/375512}, \eprint{astro-ph/0211331}

\bibitem[{{Aguerri} et~al(2001){Aguerri}, {Balcells}, and
  {Peletier}}]{AguBalPel01}
{Aguerri} JAL, {Balcells} M, {Peletier} RF (2001) {Growth of galactic bulges by
  mergers. I. Dense satellites}. \aap 367:428--442,
  \doi{10.1051/0004-6361:20000441}, \eprint{arXiv:astro-ph/0012156}

\bibitem[{{Alves-Brito} et~al(2010){Alves-Brito}, {Mel{\'e}ndez}, {Asplund},
  {Ram{\'{\i}}rez}, and {Yong}}]{alves-brito+10}
{Alves-Brito} A, {Mel{\'e}ndez} J, {Asplund} M, {Ram{\'{\i}}rez} I, {Yong} D
  (2010) {Chemical similarities between Galactic bulge and local thick disk red
  giants: O, Na, Mg, Al, Si, Ca, and Ti}. \aap 513:A35,
  \doi{10.1051/0004-6361/200913444}, \eprint{1001.2521}

\bibitem[{{Am{\^o}res} et~al(2013){Am{\^o}res}, {L{\'o}pez-Corredoira},
  {Gonz{\'a}lez-Fern{\'a}ndez}, {Moitinho}, {Minniti}, and
  {Gurovich}}]{amores+13}
{Am{\^o}res} EB, {L{\'o}pez-Corredoira} M, {Gonz{\'a}lez-Fern{\'a}ndez} C,
  {Moitinho} A, {Minniti} D, {Gurovich} S (2013) {The long bar as seen by the
  VVV Survey. II. Star counts}. \aap 559:A11,
  \doi{10.1051/0004-6361/201219846}, \eprint{1308.6022}

\bibitem[{{Athanassoula}(1992)}]{Ath92b}
{Athanassoula} E (1992) {The existence and shapes of dust lanes in galactic
  bars}. \mnras 259:345--364

\bibitem[{{Athanassoula}(2005)}]{Ath05b}
{Athanassoula} E (2005) {On the nature of bulges in general and of box/peanut
  bulges in particular: input from N-body simulations}. \mnras 358:1477--1488,
  \doi{10.1111/j.1365-2966.2005.08872.x}, \eprint{arXiv:astro-ph/0502316}

\bibitem[{{Athanassoula}(2012)}]{athanassoula+12}
{Athanassoula} E (2012) {Manifold-driven spirals in N-body barred galaxy
  simulations}. \mnras 426:L46--L50, \doi{10.1111/j.1745-3933.2012.01320.x},
  \eprint{1207.4590}

\bibitem[{{Athanassoula} and {Beaton}(2006)}]{AthBea06}
{Athanassoula} E, {Beaton} RL (2006) {Unravelling the mystery of the M31 bar}.
  \mnras 370:1499--1512, \doi{10.1111/j.1365-2966.2006.10567.x},
  \eprint{arXiv:astro-ph/0605090}

\bibitem[{{Athanassoula} and {Bureau}(1999)}]{AthBur99}
{Athanassoula} E, {Bureau} M (1999) {Bar Diagnostics in Edge-on Spiral
  Galaxies. II. Hydrodynamical Simulations}. \apj 522:699--717,
  \doi{10.1086/307677}, \eprint{arXiv:astro-ph/9904206}

\bibitem[{{Athanassoula} et~al(2005){Athanassoula}, {Lambert}, and
  {Dehnen}}]{AthLamDeh05}
{Athanassoula} E, {Lambert} JC, {Dehnen} W (2005) {Can bars be destroyed by a
  central mass concentration?- I. Simulations}. \mnras 363:496--508,
  \doi{10.1111/j.1365-2966.2005.09445.x}, \eprint{astro-ph/0507566}

\bibitem[{{Athanassoula} et~al(2014){Athanassoula}, {Laurikainen}, {Salo}, and
  {Bosma}}]{AthLauSal14}
{Athanassoula} E, {Laurikainen} E, {Salo} H, {Bosma} A (2014) {On the nature of
  the barlens component in barred galaxies}. ArXiv e-prints \eprint{1405.6726}

\bibitem[{{Baade}(1951)}]{baade+51}
{Baade} W (1951) {Galaxies - Present Day Problems}. Publications of Michigan
  Observatory 10:7

\bibitem[{{Babusiaux} and {Gilmore}(2005)}]{babusiaux+05}
{Babusiaux} C, {Gilmore} G (2005) {The structure of the Galactic bar}. \mnras
  358:1309--1319, \doi{10.1111/j.1365-2966.2005.08828.x},
  \eprint{astro-ph/0501383}

\bibitem[{{Babusiaux} et~al(2010){Babusiaux}, {G{\'o}mez}, {Hill}, {Royer},
  {Zoccali}, {Arenou}, {Fux}, {Lecureur}, {Schultheis}, {Barbuy}, {Minniti},
  and {Ortolani}}]{babusiaux+10}
{Babusiaux} C, {G{\'o}mez} A, {Hill} V, {Royer} F, {Zoccali} M, {Arenou} F,
  {Fux} R, {Lecureur} A, {Schultheis} M, {Barbuy} B, {Minniti} D, {Ortolani} S
  (2010) {Insights on the Milky Way bulge formation from the correlations
  between kinematics and metallicity}. \aap 519:A77,
  \doi{10.1051/0004-6361/201014353}, \eprint{1005.3919}

\bibitem[{{Bekki} and {Tsujimoto}(2011)}]{bekki+11}
{Bekki} K, {Tsujimoto} T (2011) {Formation of the Galactic bulge from a
  two-component stellar disc: explaining cylindrical rotation and a vertical
  metallicity gradient}. \mnras 416:L60--L64,
  \doi{10.1111/j.1745-3933.2011.01097.x}, \eprint{1106.4363}

\bibitem[{{Benjamin} et~al(2005){Benjamin}, {Churchwell}, {Babler},
  {Indebetouw}, {Meade}, {Whitney}, {Watson}, {Wolfire}, {Wolff}, {Ignace},
  {Bania}, {Bracker}, {Clemens}, {Chomiuk}, {Cohen}, {Dickey}, {Jackson},
  {Kobulnicky}, {Mercer}, {Mathis}, {Stolovy}, and {Uzpen}}]{benjamin+05}
{Benjamin} RA, {Churchwell} E, {Babler} BL, {Indebetouw} R, {Meade} MR,
  {Whitney} BA, {Watson} C, {Wolfire} MG, {Wolff} MJ, {Ignace} R, {Bania} TM,
  {Bracker} S, {Clemens} DP, {Chomiuk} L, {Cohen} M, {Dickey} JM, {Jackson} JM,
  {Kobulnicky} HA, {Mercer} EP, {Mathis} JS, {Stolovy} SR, {Uzpen} B (2005)
  {First GLIMPSE Results on the Stellar Structure of the Galaxy}. \apjl
  630:L149--L152, \doi{10.1086/491785}, \eprint{astro-ph/0508325}

\bibitem[{{Bensby} et~al(2010){Bensby}, {Feltzing}, {Johnson}, {Gould},
  {Ad{\'e}n}, {Asplund}, {Mel{\'e}ndez}, {Gal-Yam}, {Lucatello}, {Sana},
  {Sumi}, {Miyake}, {Suzuki}, {Han}, {Bond}, and {Udalski}}]{bensby+10}
{Bensby} T, {Feltzing} S, {Johnson} JA, {Gould} A, {Ad{\'e}n} D, {Asplund} M,
  {Mel{\'e}ndez} J, {Gal-Yam} A, {Lucatello} S, {Sana} H, {Sumi} T, {Miyake} N,
  {Suzuki} D, {Han} C, {Bond} I, {Udalski} A (2010) {Chemical evolution of the
  Galactic bulge as traced by microlensed dwarf and subgiant stars. II. Ages,
  metallicities, detailed elemental abundances, and connections to the Galactic
  thick disc}. \aap 512:A41, \doi{10.1051/0004-6361/200913744},
  \eprint{0911.5076}

\bibitem[{{Bensby} et~al(2011){Bensby}, {Ad{\'e}n}, {Mel{\'e}ndez}, {Gould},
  {Feltzing}, {Asplund}, {Johnson}, {Lucatello}, {Yee}, {Ram{\'{\i}}rez},
  {Cohen}, {Thompson}, {Bond}, {Gal-Yam}, {Han}, {Sumi}, {Suzuki}, {Wada},
  {Miyake}, {Furusawa}, {Ohmori}, {Saito}, {Tristram}, and
  {Bennett}}]{bensby+11}
{Bensby} T, {Ad{\'e}n} D, {Mel{\'e}ndez} J, {Gould} A, {Feltzing} S, {Asplund}
  M, {Johnson} JA, {Lucatello} S, {Yee} JC, {Ram{\'{\i}}rez} I, {Cohen} JG,
  {Thompson} I, {Bond} IA, {Gal-Yam} A, {Han} C, {Sumi} T, {Suzuki} D, {Wada}
  K, {Miyake} N, {Furusawa} K, {Ohmori} K, {Saito} T, {Tristram} P, {Bennett} D
  (2011) {Chemical evolution of the Galactic bulge as traced by microlensed
  dwarf and subgiant stars. IV. Two bulge populations}. \aap 533:A134,
  \doi{10.1051/0004-6361/201117059}, \eprint{1107.5606}

\bibitem[{{Bensby} et~al(2013{\natexlab{a}}){Bensby}, {Yee}, {Feltzing},
  {Johnson}, {Gould}, {Cohen}, {Asplund}, {Mel{\'e}ndez}, {Lucatello}, {Han},
  {Thompson}, {Gal-Yam}, {Udalski}, {Bennett}, {Bond}, {Kohei}, {Sumi},
  {Suzuki}, {Suzuki}, {Takino}, {Tristram}, {Yamai}, and
  {Yonehara}}]{bensby+13}
{Bensby} T, {Yee} JC, {Feltzing} S, {Johnson} JA, {Gould} A, {Cohen} JG,
  {Asplund} M, {Mel{\'e}ndez} J, {Lucatello} S, {Han} C, {Thompson} I,
  {Gal-Yam} A, {Udalski} A, {Bennett} DP, {Bond} IA, {Kohei} W, {Sumi} T,
  {Suzuki} D, {Suzuki} K, {Takino} S, {Tristram} P, {Yamai} N, {Yonehara} A
  (2013{\natexlab{a}}) {Chemical evolution of the Galactic bulge as traced by
  microlensed dwarf and subgiant stars. V. Evidence for a wide age distribution
  and a complex MDF}. \aap 549:A147, \doi{10.1051/0004-6361/201220678},
  \eprint{1211.6848}

\bibitem[{{Bensby} et~al(2013{\natexlab{b}}){Bensby}, {Yee}, {Feltzing},
  {Johnson}, {Gould}, {Cohen}, {Asplund}, {Mel{\'e}ndez}, {Lucatello}, {Han},
  {Thompson}, {Gal-Yam}, {Udalski}, {Bennett}, {Bond}, {Kohei}, {Sumi},
  {Suzuki}, {Suzuki}, {Takino}, {Tristram}, {Yamai}, and
  {Yonehara}}]{BenYeeFel13}
{Bensby} T, {Yee} JC, {Feltzing} S, {Johnson} JA, {Gould} A, {Cohen} JG,
  {Asplund} M, {Mel{\'e}ndez} J, {Lucatello} S, {Han} C, {Thompson} I,
  {Gal-Yam} A, {Udalski} A, {Bennett} DP, {Bond} IA, {Kohei} W, {Sumi} T,
  {Suzuki} D, {Suzuki} K, {Takino} S, {Tristram} P, {Yamai} N, {Yonehara} A
  (2013{\natexlab{b}}) {Chemical evolution of the Galactic bulge as traced by
  microlensed dwarf and subgiant stars. V. Evidence for a wide age distribution
  and a complex MDF}. \aap 549:A147, \doi{10.1051/0004-6361/201220678},
  \eprint{1211.6848}

\bibitem[{{Binney} et~al(1997){Binney}, {Gerhard}, and {Spergel}}]{binney+97}
{Binney} J, {Gerhard} O, {Spergel} D (1997) {The photometric structure of the
  inner Galaxy}. \mnras 288:365--374, \eprint{astro-ph/9609066}

\bibitem[{{Bissantz} and {Gerhard}(2002)}]{bissantz-gerhard+02}
{Bissantz} N, {Gerhard} O (2002) {Spiral arms, bar shape and bulge microlensing
  in the Milky Way}. \mnras 330:591--608,
  \doi{10.1046/j.1365-8711.2002.05116.x}, \eprint{astro-ph/0110368}

\bibitem[{{Blitz} and {Spergel}(1991)}]{blitz+91}
{Blitz} L, {Spergel} DN (1991) {Direct evidence for a bar at the Galactic
  center}. \apj 379:631--638, \doi{10.1086/170535}

\bibitem[{{Bouch{\'e}} et~al(2013){Bouch{\'e}}, {Murphy}, {Kacprzak},
  {P{\'e}roux}, {Contini}, {Martin}, and {Dessauges-Zavadsky}}]{BouMurKac13}
{Bouch{\'e}} N, {Murphy} MT, {Kacprzak} GG, {P{\'e}roux} C, {Contini} T,
  {Martin} CL, {Dessauges-Zavadsky} M (2013) {Signatures of Cool Gas Fueling a
  Star-Forming Galaxy at Redshift 2.3}. Science 341:50--53,
  \doi{10.1126/science.1234209}, \eprint{1306.0134}

\bibitem[{{Bournaud} and {Combes}(2002)}]{BouCom02}
{Bournaud} F, {Combes} F (2002) {Gas accretion on spiral galaxies: Bar
  formation and renewal}. \aap 392:83--102, \doi{10.1051/0004-6361:20020920},
  \eprint{arXiv:astro-ph/0206273}

\bibitem[{{Bournaud} et~al(2005){Bournaud}, {Combes}, and
  {Semelin}}]{BouComSem05}
{Bournaud} F, {Combes} F, {Semelin} B (2005) {The lifetime of galactic bars:
  central mass concentrations and gravity torques}. \mnras 364:L18--L22,
  \doi{10.1111/j.1745-3933.2005.00096.x}, \eprint{arXiv:astro-ph/0509126}

\bibitem[{{Burbidge} and {Burbidge}(1959)}]{BurBur59}
{Burbidge} EM, {Burbidge} GR (1959) {Three Unusual so Galaxies.} \apj 130:20,
  \doi{10.1086/146691}

\bibitem[{{Bureau} and {Athanassoula}(1999)}]{BurAth99}
{Bureau} M, {Athanassoula} E (1999) {Bar Diagnostics in Edge-on Spiral
  Galaxies. I. The Periodic Orbits Approach}. \apj 522:686--698,
  \doi{10.1086/307675}, \eprint{arXiv:astro-ph/9903061}

\bibitem[{{Bureau} and {Freeman}(1999)}]{BurFre99}
{Bureau} M, {Freeman} KC (1999) {The Nature of Boxy/Peanut-Shaped Bulges in
  Spiral Galaxies}. \aj 118:126--138, \doi{10.1086/300922},
  \eprint{arXiv:astro-ph/9904015}

\bibitem[{{Cabrera-Lavers} et~al(2007){Cabrera-Lavers}, {Hammersley},
  {Gonz{\'a}lez-Fern{\'a}ndez}, {L{\'o}pez-Corredoira}, {Garz{\'o}n}, and
  {Mahoney}}]{cabrera-lavers+07}
{Cabrera-Lavers} A, {Hammersley} PL, {Gonz{\'a}lez-Fern{\'a}ndez} C,
  {L{\'o}pez-Corredoira} M, {Garz{\'o}n} F, {Mahoney} TJ (2007) {Tracing the
  long bar with red-clump giants}. \aap 465:825--838,
  \doi{10.1051/0004-6361:20066185}, \eprint{astro-ph/0702109}

\bibitem[{{Cao} et~al(2013){Cao}, {Mao}, {Nataf}, {Rattenbury}, and
  {Gould}}]{cao+13}
{Cao} L, {Mao} S, {Nataf} D, {Rattenbury} NJ, {Gould} A (2013) {A new
  photometric model of the Galactic bar using red clump giants}. \mnras
  434:595--605, \doi{10.1093/mnras/stt1045}, \eprint{1303.6430}

\bibitem[{{Carollo} et~al(1997){Carollo}, {Stiavelli}, {de Zeeuw}, and
  {Mack}}]{CarStideZ97}
{Carollo} CM, {Stiavelli} M, {de Zeeuw} PT, {Mack} J (1997) {Spiral Galaxies
  with WFPC2.I.Nuclear Morphology, Bulges, Star Clusters, and Surface
  Brightness Profiles}. \aj 114:2366--+, \doi{10.1086/118654}

\bibitem[{{Churchwell} et~al(2009){Churchwell}, {Babler}, {Meade}, {Whitney},
  {Benjamin}, {Indebetouw}, {Cyganowski}, {Robitaille}, {Povich}, {Watson}, and
  {Bracker}}]{churchwell+09}
{Churchwell} E, {Babler} BL, {Meade} MR, {Whitney} BA, {Benjamin} R,
  {Indebetouw} R, {Cyganowski} C, {Robitaille} TP, {Povich} M, {Watson} C,
  {Bracker} S (2009) {The Spitzer/GLIMPSE Surveys: A New View of the Milky
  Way}. \pasp 121:213--230, \doi{10.1086/597811}

\bibitem[{{Clarkson} et~al(2008){Clarkson}, {Sahu}, {Anderson}, {Smith},
  {Brown}, {Rich}, {Casertano}, {Bond}, {Livio}, {Minniti}, {Panagia},
  {Renzini}, {Valenti}, and {Zoccali}}]{clarkson+08}
{Clarkson} W, {Sahu} K, {Anderson} J, {Smith} TE, {Brown} TM, {Rich} RM,
  {Casertano} S, {Bond} HE, {Livio} M, {Minniti} D, {Panagia} N, {Renzini} A,
  {Valenti} J, {Zoccali} M (2008) {Stellar Proper Motions in the Galactic Bulge
  from Deep Hubble Space Telescope ACS WFC Photometry}. \apj 684:1110--1142,
  \doi{10.1086/590378}, \eprint{0809.1682}

\bibitem[{{Clarkson} et~al(2011){Clarkson}, {Sahu}, {Anderson}, {Rich},
  {Smith}, {Brown}, {Bond}, {Livio}, {Minniti}, {Renzini}, and
  {Zoccali}}]{clarkson+11}
{Clarkson} WI, {Sahu} KC, {Anderson} J, {Rich} RM, {Smith} TE, {Brown} TM,
  {Bond} HE, {Livio} M, {Minniti} D, {Renzini} A, {Zoccali} M (2011) {The First
  Detection of Blue Straggler Stars in the Milky Way Bulge}. \apj 735:37,
  \doi{10.1088/0004-637X/735/1/37}, \eprint{1105.4176}

\bibitem[{{Coelho} and {Gadotti}(2011)}]{CoeGad11}
{Coelho} P, {Gadotti} DA (2011) {Bars Rejuvenating Bulges? Evidence from
  Stellar Population Analysis}. \apjl 743:L13,
  \doi{10.1088/2041-8205/743/1/L13}, \eprint{1111.1736}

\bibitem[{{Combes} and {Sanders}(1981)}]{ComSan81}
{Combes} F, {Sanders} RH (1981) {Formation and properties of persisting stellar
  bars}. \aap 96:164--173

\bibitem[{{Comer{\'o}n} et~al(2011){Comer{\'o}n}, {Elmegreen}, {Knapen},
  {Salo}, {Laurikainen}, {Laine}, {Athanassoula}, {Bosma}, {Sheth}, {Regan},
  {Hinz}, {Gil de Paz}, {Men{\'e}ndez-Delmestre}, {Mizusawa},
  {Mu{\~n}oz-Mateos}, {Seibert}, {Kim}, {Elmegreen}, {Gadotti}, {Ho},
  {Holwerda}, {Lappalainen}, {Schinnerer}, and {Skibba}}]{ComElmKna11}
{Comer{\'o}n} S, {Elmegreen} BG, {Knapen} JH, {Salo} H, {Laurikainen} E,
  {Laine} J, {Athanassoula} E, {Bosma} A, {Sheth} K, {Regan} MW, {Hinz} JL,
  {Gil de Paz} A, {Men{\'e}ndez-Delmestre} K, {Mizusawa} T, {Mu{\~n}oz-Mateos}
  JC, {Seibert} M, {Kim} T, {Elmegreen} DM, {Gadotti} DA, {Ho} LC, {Holwerda}
  BW, {Lappalainen} J, {Schinnerer} E, {Skibba} R (2011) {Thick Disks of
  Edge-on Galaxies Seen through the Spitzer Survey of Stellar Structure in
  Galaxies (S$^{4}$G): Lair of Missing Baryons?} \apj 741:28,
  \doi{10.1088/0004-637X/741/1/28}, \eprint{1108.0037}

\bibitem[{{Comer{\'o}n} et~al(2014){Comer{\'o}n}, {Elmegreen}, {Salo},
  {Laurikainen}, {Holwerda}, and {Knapen}}]{ComElmSal14}
{Comer{\'o}n} S, {Elmegreen} BG, {Salo} H, {Laurikainen} E, {Holwerda} BW,
  {Knapen} JH (2014) {Evidence for the concurrent growth of thick discs and
  central mass concentrations from S$^{4}$G imaging}. \aap 571:A58,
  \doi{10.1051/0004-6361/201424412}, \eprint{1409.0466}

\bibitem[{{Courteau} et~al(2011){Courteau}, {Widrow}, {McDonald},
  {Guhathakurta}, {Gilbert}, {Zhu}, {Beaton}, and {Majewski}}]{CouWidMcD11}
{Courteau} S, {Widrow} LM, {McDonald} M, {Guhathakurta} P, {Gilbert} KM, {Zhu}
  Y, {Beaton} RL, {Majewski} SR (2011) {The Luminosity Profile and Structural
  Parameters of the Andromeda Galaxy}. \apj 739:20,
  \doi{10.1088/0004-637X/739/1/20}, \eprint{1106.3564}

\bibitem[{{Cunha} and {Smith}(2006)}]{cunha-smith+06}
{Cunha} K, {Smith} VV (2006) {Chemical Evolution of the Galactic Bulge as
  Derived from High-Resolution Infrared Spectroscopy of K and M Red Giants}.
  \apj 651:491--501, \doi{10.1086/507673}, \eprint{astro-ph/0607393}

\bibitem[{{de Souza} and {Dos Anjos}(1987)}]{deSDos87}
{de Souza} RE, {Dos Anjos} S (1987) {Box-shaped galaxies - A complete list}.
  \aaps 70:465--480

\bibitem[{{de Vaucouleurs}(1964)}]{deVaucouleurs+64}
{de Vaucouleurs} G (1964) {Interpretation of velocity distribution of the inner
  regions of the Galaxy}. In: {Kerr} FJ (ed) The Galaxy and the Magellanic
  Clouds, IAU Symposium, vol~20, p 195

\bibitem[{{de Vaucouleurs}(1974)}]{deV74}
{de Vaucouleurs} G (1974) {Structures of Central Bulges and Nuclei of
  Galaxies}. In: {Shakeshaft} JR (ed) The Formation and Dynamics of Galaxies,
  IAU Symposium, vol~58, p 335

\bibitem[{{Debattista} et~al(2005){Debattista}, {Carollo}, {Mayer}, and
  {Moore}}]{debattista+05}
{Debattista} VP, {Carollo} CM, {Mayer} L, {Moore} B (2005) {The Kinematic
  Signature of Face-On Peanut-shaped Bulges}. \apj 628:678--694,
  \doi{10.1086/431292}, \eprint{astro-ph/0504530}

\bibitem[{{D{\'e}k{\'a}ny} et~al(2013){D{\'e}k{\'a}ny}, {Minniti}, {Catelan},
  {Zoccali}, {Saito}, {Hempel}, and {Gonzalez}}]{dekany+13}
{D{\'e}k{\'a}ny} I, {Minniti} D, {Catelan} M, {Zoccali} M, {Saito} RK, {Hempel}
  M, {Gonzalez} OA (2013) {VVV Survey Near-infrared Photometry of Known Bulge
  RR Lyrae Stars: The Distance to the Galactic Center and Absence of a Barred
  Distribution of the Metal-poor Population}. \apjl 776:L19,
  \doi{10.1088/2041-8205/776/2/L19}, \eprint{1309.5933}

\bibitem[{{Di Matteo} et~al(2014){Di Matteo}, {Gomez}, {Haywood}, {Combes},
  {Lehnert}, {Ness}, {Snaith}, {Katz}, and {Semelin}}]{dimatteo+14}
{Di Matteo} P, {Gomez} A, {Haywood} M, {Combes} F, {Lehnert} MD, {Ness} M,
  {Snaith} ON, {Katz} D, {Semelin} B (2014) {Why the Milky Way's bulge is not
  only a bar formed from a cold thin disk}. ArXiv e-prints \eprint{1411.1416}

\bibitem[{{Dwek} et~al(1995){Dwek}, {Arendt}, {Hauser}, {Kelsall}, {Lisse},
  {Moseley}, {Silverberg}, {Sodroski}, and {Weiland}}]{dwek+95}
{Dwek} E, {Arendt} RG, {Hauser} MG, {Kelsall} T, {Lisse} CM, {Moseley} SH,
  {Silverberg} RF, {Sodroski} TJ, {Weiland} JL (1995) {Morphology,
  near-infrared luminosity, and mass of the Galactic bulge from COBE DIRBE
  observations}. \apj 445:716--730, \doi{10.1086/175734}

\bibitem[{{Eggen} et~al(1962){Eggen}, {Lynden-Bell}, and
  {Sandage}}]{EggLynSan62}
{Eggen} OJ, {Lynden-Bell} D, {Sandage} AR (1962) {Evidence from the motions of
  old stars that the Galaxy collapsed.} \apj 136:748--+

\bibitem[{{Ellison} et~al(2011){Ellison}, {Nair}, {Patton}, {Scudder},
  {Mendel}, and {Simard}}]{EllNaiPat2011}
{Ellison} SL, {Nair} P, {Patton} DR, {Scudder} JM, {Mendel} JT, {Simard} L
  (2011) {The impact of gas inflows on star formation rates and metallicities
  in barred galaxies}. \mnras 416:2182--2192,
  \doi{10.1111/j.1365-2966.2011.19195.x}, \eprint{1106.1177}

\bibitem[{{Elmegreen}(1999)}]{elmegreen+99}
{Elmegreen} BG (1999) {Galactic Bulge Formation as a Maximum Intensity
  Starburst}. \apj 517:103--107, \doi{10.1086/307200},
  \eprint{astro-ph/9901025}

\bibitem[{{Elmegreen} et~al(2008){Elmegreen}, {Bournaud}, and
  {Elmegreen}}]{elmegreen+08}
{Elmegreen} BG, {Bournaud} F, {Elmegreen} DM (2008) {Bulge Formation by the
  Coalescence of Giant Clumps in Primordial Disk Galaxies}. \apj 688:67--77,
  \doi{10.1086/592190}, \eprint{0808.0716}

\bibitem[{{Emsellem} et~al(2014){Emsellem}, {Renaud}, {Bournaud}, {Elmegreen},
  {Combes}, and {Gabor}}]{EmsRenBou14}
{Emsellem} E, {Renaud} F, {Bournaud} F, {Elmegreen} B, {Combes} F, {Gabor} J
  (2014) {The interplay between a galactic bar and a supermassive black hole:
  nuclear fueling in a sub-parsec resolution galaxy simulation}. ArXiv e-prints
  \eprint{1410.6479}

\bibitem[{{Erwin} and {Debattista}(2013)}]{ErwDeb13}
{Erwin} P, {Debattista} VP (2013) {Peanuts at an angle: detecting and measuring
  the three-dimensional structure of bars in moderately inclined galaxies}.
  \mnras 431:3060--3086, \doi{10.1093/mnras/stt385}, \eprint{1301.0638}

\bibitem[{{Erwin} et~al(2014){Erwin}, {Saglia}, {Fabricius}, {Thomas}, {Nowak},
  {Rusli}, {Bender}, {Vega Beltran}, and {Beckman}}]{ErwSagFab14}
{Erwin} P, {Saglia} RP, {Fabricius} M, {Thomas} J, {Nowak} N, {Rusli} S,
  {Bender} R, {Vega Beltran} JC, {Beckman} JE (2014) {Composite Bulges: The
  Coexistence of Classical Bulges and Disky Pseudobulges in S0 and Spiral
  Galaxies}. ArXiv e-prints \eprint{1411.2599}

\bibitem[{{Fisher}(2006)}]{Fis06}
{Fisher} DB (2006) {Central Star Formation and PAH Profiles in Pseudobulges and
  Classical Bulges}. \apjl 642:L17--L20, \doi{10.1086/504351},
  \eprint{arXiv:astro-ph/0603455}

\bibitem[{{Fisher} and {Drory}(2010)}]{FisDro10}
{Fisher} DB, {Drory} N (2010) {Bulges of Nearby Galaxies with Spitzer: Scaling
  Relations in Pseudobulges and Classical Bulges}. \apj 716:942--969,
  \doi{10.1088/0004-637X/716/2/942}, \eprint{1004.5393}

\bibitem[{{Freeman} et~al(2013){Freeman}, {Ness}, {Wylie-de-Boer},
  {Athanassoula}, {Bland-Hawthorn}, {Asplund}, {Lewis}, {Yong}, {Lane}, {Kiss},
  and {Ibata}}]{freeman-argos+13}
{Freeman} K, {Ness} M, {Wylie-de-Boer} E, {Athanassoula} E, {Bland-Hawthorn} J,
  {Asplund} M, {Lewis} G, {Yong} D, {Lane} R, {Kiss} L, {Ibata} R (2013) {ARGOS
  - II. The Galactic bulge survey}. \mnras 428:3660--3670,
  \doi{10.1093/mnras/sts305}, \eprint{1212.1541}

\bibitem[{{Frogel} and {Whitford}(1987)}]{frogel+87}
{Frogel} JA, {Whitford} AE (1987) {M giants in Baade's window - Infrared
  colors, luminosities, and implications for the stellar content of E and S0
  galaxies}. \apj 320:199--237, \doi{10.1086/165535}

\bibitem[{{Fulbright} et~al(2006){Fulbright}, {McWilliam}, and
  {Rich}}]{fulbright-mcwilliam-rich+06}
{Fulbright} JP, {McWilliam} A, {Rich} RM (2006) {Abundances of Baade's Window
  Giants from Keck HIRES Spectra. I. Stellar Parameters and [Fe/H] Values}.
  \apj 636:821--841, \doi{10.1086/498205}, \eprint{astro-ph/0510408}

\bibitem[{{Fulbright} et~al(2007){Fulbright}, {McWilliam}, and
  {Rich}}]{fulbright-mcwilliam-rich+07}
{Fulbright} JP, {McWilliam} A, {Rich} RM (2007) {Abundances of Baade's Window
  Giants from Keck HIRES Spectra. II. The Alpha and Light Odd Elements}. \apj
  661:1152--1179, \doi{10.1086/513710}, \eprint{astro-ph/0609087}

\bibitem[{{Gadotti}(2008)}]{Gad08}
{Gadotti} DA (2008) {Image decomposition of barred galaxies and AGN hosts}.
  \mnras 384:420--439, \doi{10.1111/j.1365-2966.2007.12723.x},
  \eprint{arXiv:0708.3870}

\bibitem[{{Gadotti}(2009)}]{Gad09b}
{Gadotti} DA (2009) {Structural properties of pseudo-bulges, classical bulges
  and elliptical galaxies: a Sloan Digital Sky Survey perspective}. \mnras
  393:1531--1552, \doi{10.1111/j.1365-2966.2008.14257.x}, \eprint{0810.1953}

\bibitem[{{Gadotti}(2012)}]{Gad12}
{Gadotti} DA (2012) {Galaxy Bulges and Elliptical Galaxies. Lecture Notes}.
  {Memorie della Societa Astronomica Italiana} in press

\bibitem[{{Gadotti} and {dos Anjos}(2001)}]{GadDos01}
{Gadotti} DA, {dos Anjos} S (2001) {Homogenization of the Stellar Population
  along Late-Type Spiral Galaxies}. \aj 122:1298--1318, \doi{10.1086/322126},
  \eprint{arXiv:astro-ph/0106303}

\bibitem[{{Gardner} et~al(2014){Gardner}, {Debattista}, {Robin}, {V{\'a}squez},
  and {Zoccali}}]{gardner+14}
{Gardner} E, {Debattista} VP, {Robin} AC, {V{\'a}squez} S, {Zoccali} M (2014)
  {N-body simulation insights into the X-shaped bulge of the Milky Way:
  kinematics and distance to the Galactic Centre}. \mnras 438:3275--3290,
  \doi{10.1093/mnras/stt2430}, \eprint{1306.4694}

\bibitem[{{Garz{\'o}n} and {L{\'o}pez-Corredoira}(2014)}]{garzon+14}
{Garz{\'o}n} F, {L{\'o}pez-Corredoira} M (2014) {Dynamical evolution of two
  associated galactic bars}. Astronomische Nachrichten 335:865,
  \doi{10.1002/asna.201412120}, \eprint{1409.1916}

\bibitem[{{Gerhard} and {Martinez-Valpuesta}(2012)}]{gerhard+12}
{Gerhard} O, {Martinez-Valpuesta} I (2012) {The Inner Galactic Bulge: Evidence
  for a Nuclear Bar?} \apjl 744:L8, \doi{10.1088/2041-8205/744/1/L8},
  \eprint{1112.0179}

\bibitem[{{Gonzalez} et~al(2011){Gonzalez}, {Rejkuba}, {Zoccali}, {Hill},
  {Battaglia}, {Babusiaux}, {Minniti}, {Barbuy}, {Alves-Brito}, {Renzini},
  {Gomez}, and {Ortolani}}]{gonzalez+11}
{Gonzalez} OA, {Rejkuba} M, {Zoccali} M, {Hill} V, {Battaglia} G, {Babusiaux}
  C, {Minniti} D, {Barbuy} B, {Alves-Brito} A, {Renzini} A, {Gomez} A,
  {Ortolani} S (2011) {Alpha element abundances and gradients in the Milky Way
  bulge from FLAMES-GIRAFFE spectra of 650 K giants}. \aap 530:A54,
  \doi{10.1051/0004-6361/201116548}, \eprint{1103.6104}

\bibitem[{{Gonzalez} et~al(2012){Gonzalez}, {Rejkuba}, {Zoccali}, {Valenti},
  {Minniti}, {Schultheis}, {Tobar}, and {Chen}}]{gonzalez+12}
{Gonzalez} OA, {Rejkuba} M, {Zoccali} M, {Valenti} E, {Minniti} D, {Schultheis}
  M, {Tobar} R, {Chen} B (2012) {Reddening and metallicity maps of the Milky
  Way bulge from VVV and 2MASS. II. The complete high resolution extinction map
  and implications for Galactic bulge studies}. \aap 543:A13,
  \doi{10.1051/0004-6361/201219222}, \eprint{1204.4004}

\bibitem[{{Gonzalez} et~al(2013){Gonzalez}, {Rejkuba}, {Zoccali}, {Valent},
  {Minniti}, and {Tobar}}]{gonzalez+13}
{Gonzalez} OA, {Rejkuba} M, {Zoccali} M, {Valent} E, {Minniti} D, {Tobar} R
  (2013) {Reddening and metallicity maps of the Milky Way bulge from VVV and
  2MASS. III. The first global photometric metallicity map of the Galactic
  bulge}. \aap 552:A110, \doi{10.1051/0004-6361/201220842}, \eprint{1302.0243}

\bibitem[{{Hammersley} et~al(2000){Hammersley}, {Garz{\'o}n}, {Mahoney},
  {L{\'o}pez-Corredoira}, and {Torres}}]{hammersley+00}
{Hammersley} PL, {Garz{\'o}n} F, {Mahoney} TJ, {L{\'o}pez-Corredoira} M,
  {Torres} MAP (2000) {Detection of the old stellar component of the major
  Galactic bar}. \mnras 317:L45--L49, \doi{10.1046/j.1365-8711.2000.03858.x},
  \eprint{astro-ph/0007232}

\bibitem[{{Hill} et~al(2011){Hill}, {Lecureur}, {G{\'o}mez}, {Zoccali},
  {Schultheis}, {Babusiaux}, {Royer}, {Barbuy}, {Arenou}, {Minniti}, and
  {Ortolani}}]{hill+11}
{Hill} V, {Lecureur} A, {G{\'o}mez} A, {Zoccali} M, {Schultheis} M, {Babusiaux}
  C, {Royer} F, {Barbuy} B, {Arenou} F, {Minniti} D, {Ortolani} S (2011) {The
  metallicity distribution of bulge clump giants in Baade's window}. \aap
  534:A80, \doi{10.1051/0004-6361/200913757}, \eprint{1107.5199}

\bibitem[{{Hopkins} et~al(2010){Hopkins}, {Bundy}, {Croton}, {Hernquist},
  {Keres}, {Khochfar}, {Stewart}, {Wetzel}, and {Younger}}]{HopBunCro10}
{Hopkins} PF, {Bundy} K, {Croton} D, {Hernquist} L, {Keres} D, {Khochfar} S,
  {Stewart} K, {Wetzel} A, {Younger} JD (2010) {Mergers and Bulge Formation in
  {$\Lambda$}CDM: Which Mergers Matter?} \apj 715:202--229,
  \doi{10.1088/0004-637X/715/1/202}, \eprint{0906.5357}

\bibitem[{{Howard} et~al(2008){Howard}, {Rich}, {Reitzel}, {Koch}, {De
  Propris}, and {Zhao}}]{howard+08}
{Howard} CD, {Rich} RM, {Reitzel} DB, {Koch} A, {De Propris} R, {Zhao} H (2008)
  {The Bulge Radial Velocity Assay (BRAVA). I. Sample Selection and a Rotation
  Curve}. \apj 688:1060--1077, \doi{10.1086/592106}, \eprint{0807.3967}

\bibitem[{{Jarvis}(1986)}]{Jar86}
{Jarvis} BJ (1986) {A search for box- or peanut-shaped bulges}. \aj 91:65--69,
  \doi{10.1086/113980}

\bibitem[{{Johnson} et~al(2011){Johnson}, {Rich}, {Fulbright}, {Valenti}, and
  {McWilliam}}]{johnson+11}
{Johnson} CI, {Rich} RM, {Fulbright} JP, {Valenti} E, {McWilliam} A (2011)
  {Alpha Enhancement and the Metallicity Distribution Function of Plaut's
  Window}. \apj 732:108, \doi{10.1088/0004-637X/732/2/108}, \eprint{1103.2143}

\bibitem[{{Johnson} et~al(2013){Johnson}, {Rich}, {Kobayashi}, {Kunder},
  {Pilachowski}, {Koch}, and {de Propris}}]{johnson+13}
{Johnson} CI, {Rich} RM, {Kobayashi} C, {Kunder} A, {Pilachowski} CA, {Koch} A,
  {de Propris} R (2013) {Metallicity Distribution Functions, Radial Velocities,
  and Alpha Element Abundances in Three Off-axis Bulge Fields}. \apj 765:157,
  \doi{10.1088/0004-637X/765/2/157}, \eprint{1302.3679}

\bibitem[{{Johnson} et~al(2014){Johnson}, {Rich}, {Kobayashi}, {Kunder}, and
  {Koch}}]{johnson+14}
{Johnson} CI, {Rich} RM, {Kobayashi} C, {Kunder} A, {Koch} A (2014) {Light,
  Alpha, and Fe-peak Element Abundances in the Galactic Bulge}. \aj 148:67,
  \doi{10.1088/0004-6256/148/4/67}, \eprint{1407.2282}

\bibitem[{{Kormendy} and {Barentine}(2010)}]{KorBar10}
{Kormendy} J, {Barentine} JC (2010) {Detection of a Pseudobulge Hidden Inside
  the ''Box-shaped Bulge'' of NGC 4565}. \apjl 715:L176--L179,
  \doi{10.1088/2041-8205/715/2/L176}, \eprint{1005.1647}

\bibitem[{{Kormendy} and {Kennicutt}(2004)}]{KorKen04}
{Kormendy} J, {Kennicutt} RC Jr (2004) {Secular Evolution and the Formation of
  Pseudobulges in Disk Galaxies}. \araa 42:603--683,
  \eprint{arXiv:astro-ph/0407343}

\bibitem[{{Kraljic} et~al(2012){Kraljic}, {Bournaud}, and
  {Martig}}]{KraBouMar12}
{Kraljic} K, {Bournaud} F, {Martig} M (2012) {The Two-phase Formation History
  of Spiral Galaxies Traced by the Cosmic Evolution of the Bar Fraction}. \apj
  757:60, \doi{10.1088/0004-637X/757/1/60}, \eprint{1207.0351}

\bibitem[{{Kuijken} and {Merrifield}(1995)}]{KuiMer95}
{Kuijken} K, {Merrifield} MR (1995) {Establishing the connection between
  peanut-shaped bulges and galactic bars}. \apjl 443:L13--L16,
  \doi{10.1086/187824}, \eprint{arXiv:astro-ph/9501114}

\bibitem[{{Kuijken} and {Rich}(2002)}]{kuijken-rich+02}
{Kuijken} K, {Rich} RM (2002) {Hubble Space Telescope WFPC2 Proper Motions in
  Two Bulge Fields: Kinematics and Stellar Population of the Galactic Bulge}.
  \aj 124:2054--2066, \doi{10.1086/342540}, \eprint{astro-ph/0207116}

\bibitem[{{Kunder} et~al(2012){Kunder}, {Koch}, {Rich}, {de Propris}, {Howard},
  {Stubbs}, {Johnson}, {Shen}, {Wang}, {Robin}, {Kormendy}, {Soto},
  {Frinchaboy}, {Reitzel}, {Zhao}, and {Origlia}}]{kunder+12}
{Kunder} A, {Koch} A, {Rich} RM, {de Propris} R, {Howard} CD, {Stubbs} SA,
  {Johnson} CI, {Shen} J, {Wang} Y, {Robin} AC, {Kormendy} J, {Soto} M,
  {Frinchaboy} P, {Reitzel} DB, {Zhao} H, {Origlia} L (2012) {The Bulge Radial
  Velocity Assay (BRAVA). II. Complete Sample and Data Release}. \aj 143:57,
  \doi{10.1088/0004-6256/143/3/57}, \eprint{1112.1955}

\bibitem[{{Laurikainen} et~al(2005){Laurikainen}, {Salo}, and
  {Buta}}]{LauSalBut05}
{Laurikainen} E, {Salo} H, {Buta} R (2005) {Multicomponent decompositions for a
  sample of S0 galaxies}. \mnras 362:1319--1347,
  \doi{10.1111/j.1365-2966.2005.09404.x}, \eprint{arXiv:astro-ph/0508097}

\bibitem[{{Laurikainen} et~al(2007){Laurikainen}, {Salo}, {Buta}, and
  {Knapen}}]{LauSalBut07}
{Laurikainen} E, {Salo} H, {Buta} R, {Knapen} JH (2007) {Properties of bars and
  bulges in the Hubble sequence}. \mnras 381:401--417,
  \doi{10.1111/j.1365-2966.2007.12299.x}, \eprint{arXiv:astro-ph/0702434}

\bibitem[{{Laurikainen} et~al(2011){Laurikainen}, {Salo}, {Buta}, and
  {Knapen}}]{LauSalBut11}
{Laurikainen} E, {Salo} H, {Buta} R, {Knapen} JH (2011) {Near-infrared atlas of
  S0-Sa galaxies (NIRS0S)}. \mnras 418:1452--1490,
  \doi{10.1111/j.1365-2966.2011.19283.x}, \eprint{1110.1996}

\bibitem[{{Laurikainen} et~al(2014){Laurikainen}, {Salo}, {Athanassoula},
  {Bosma}, and {Herrera-Endoqui}}]{LauSalAth14}
{Laurikainen} E, {Salo} H, {Athanassoula} E, {Bosma} A, {Herrera-Endoqui} M
  (2014) {Milky Way mass galaxies with X-shaped bulges are not rare in the
  local Universe}. \mnras 444:L80--L84, \doi{10.1093/mnrasl/slu118},
  \eprint{1406.1418}

\bibitem[{{Lecureur} et~al(2007){Lecureur}, {Hill}, {Zoccali}, {Barbuy},
  {G{\'o}mez}, {Minniti}, {Ortolani}, and {Renzini}}]{lecureur+07}
{Lecureur} A, {Hill} V, {Zoccali} M, {Barbuy} B, {G{\'o}mez} A, {Minniti} D,
  {Ortolani} S, {Renzini} A (2007) {Oxygen, sodium, magnesium, and aluminium as
  tracers of the galactic bulge formation}. \aap 465:799--814,
  \doi{10.1051/0004-6361:20066036}, \eprint{astro-ph/0610346}

\bibitem[{{Liszt} and {Burton}(1980)}]{liszt+80}
{Liszt} HS, {Burton} WB (1980) {The gas distribution in the central region of
  the Galaxy. III - A barlike model of the inner-Galaxy gas based on improved H
  I data}. \apj 236:779--797, \doi{10.1086/157803}

\bibitem[{{L{\'o}pez-Corredoira} et~al(2007){L{\'o}pez-Corredoira},
  {Cabrera-Lavers}, {Mahoney}, {Hammersley}, {Garz{\'o}n}, and
  {Gonz{\'a}lez-Fern{\'a}ndez}}]{lopez-corredoira+07}
{L{\'o}pez-Corredoira} M, {Cabrera-Lavers} A, {Mahoney} TJ, {Hammersley} PL,
  {Garz{\'o}n} F, {Gonz{\'a}lez-Fern{\'a}ndez} C (2007) {The Long Bar in the
  Milky Way: Corroboration of an Old Hypothesis}. \aj 133:154--161,
  \doi{10.1086/509605}, \eprint{astro-ph/0606201}

\bibitem[{{L{\"u}tticke} et~al(2000){L{\"u}tticke}, {Dettmar}, and
  {Pohlen}}]{LueDetPoh00}
{L{\"u}tticke} R, {Dettmar} RJ, {Pohlen} M (2000) {Box- and peanut-shaped
  bulges. I. Statistics}. \aaps 145:405--414, \doi{10.1051/aas:2000354},
  \eprint{arXiv:astro-ph/0006359}

\bibitem[{{Martinez-Valpuesta} and {Gerhard}(2011)}]{martinez-valpuesta+11}
{Martinez-Valpuesta} I, {Gerhard} O (2011) {Unifying A Boxy Bulge and Planar
  Long Bar in the Milky Way}. \apjl 734:L20, \doi{10.1088/2041-8205/734/1/L20},
  \eprint{1105.0928}

\bibitem[{{Martinez-Valpuesta} and {Gerhard}(2013)}]{martinez-valpuesta+13}
{Martinez-Valpuesta} I, {Gerhard} O (2013) {Metallicity Gradients Through Disk
  Instability: A Simple Model for the Milky Way's Boxy Bulge}. \apjl 766:L3,
  \doi{10.1088/2041-8205/766/1/L3}, \eprint{1302.1613}

\bibitem[{{Martinez-Valpuesta} et~al(2006){Martinez-Valpuesta}, {Shlosman}, and
  {Heller}}]{MarShlHel06}
{Martinez-Valpuesta} I, {Shlosman} I, {Heller} C (2006) {Evolution of Stellar
  Bars in Live Axisymmetric Halos: Recurrent Buckling and Secular Growth}. \apj
  637:214--226, \doi{10.1086/498338}, \eprint{arXiv:astro-ph/0507219}

\bibitem[{{McWilliam} and {Rich}(1994)}]{McWRic94}
{McWilliam} A, {Rich} RM (1994) {The first detailed abundance analysis of
  Galactic bulge K giants in Baade's window}. \apjs 91:749--791,
  \doi{10.1086/191954}

\bibitem[{{McWilliam} and {Zoccali}(2010)}]{mcwilliam-zoccali+10}
{McWilliam} A, {Zoccali} M (2010) {Two Red Clumps and the X-shaped Milky Way
  Bulge}. \apj 724:1491--1502, \doi{10.1088/0004-637X/724/2/1491},
  \eprint{1008.0519}

\bibitem[{{McWilliam} et~al(2010){McWilliam}, {Fulbright}, and
  {Rich}}]{mcwilliam-fulbright-rich+10}
{McWilliam} A, {Fulbright} J, {Rich} RM (2010) {Chemical Composition of the
  Galactic Bulge in Baade's Window}. In: {Cunha} K, {Spite} M, {Barbuy} B (eds)
  IAU Symposium, IAU Symposium, vol 265, pp 279--284,
  \doi{10.1017/S1743921310000748}

\bibitem[{{Mel{\'e}ndez} et~al(2008){Mel{\'e}ndez}, {Asplund}, {Alves-Brito},
  {Cunha}, {Barbuy}, {Bessell}, {Chiappini}, {Freeman}, {Ram{\'{\i}}rez},
  {Smith}, and {Yong}}]{melendez+08}
{Mel{\'e}ndez} J, {Asplund} M, {Alves-Brito} A, {Cunha} K, {Barbuy} B,
  {Bessell} MS, {Chiappini} C, {Freeman} KC, {Ram{\'{\i}}rez} I, {Smith} VV,
  {Yong} D (2008) {Chemical similarities between Galactic bulge and local thick
  disk red giant stars}. \aap 484:L21--L25, \doi{10.1051/0004-6361:200809398},
  \eprint{0804.4124}

\bibitem[{{M{\'e}ndez-Abreu} et~al(2014){M{\'e}ndez-Abreu}, {Debattista},
  {Corsini}, and {Aguerri}}]{MenDebCor14}
{M{\'e}ndez-Abreu} J, {Debattista} VP, {Corsini} EM, {Aguerri} JAL (2014)
  {Secular- and merger-built bulges in barred galaxies}. \aap 572:A25,
  \doi{10.1051/0004-6361/201423955}, \eprint{1409.2876}

\bibitem[{{Merrifield} and {Kuijken}(1999)}]{MerKui99}
{Merrifield} MR, {Kuijken} K (1999) {Hidden bars and boxy bulges}. \aap
  345:L47--L50, \eprint{arXiv:astro-ph/9904158}

\bibitem[{{Minniti}(1996)}]{minniti+96}
{Minniti} D (1996) {Kinematics of Bulge Giants in F588}. \apj 459:579,
  \doi{10.1086/176923}

\bibitem[{{Minniti} and {Zoccali}(2008)}]{minniti-zoccali+08}
{Minniti} D, {Zoccali} M (2008) {The Galactic bulge: a review}. In: {Bureau} M,
  {Athanassoula} E, {Barbuy} B (eds) IAU Symposium, IAU Symposium, vol 245, pp
  323--332, \doi{10.1017/S1743921308018048}, \eprint{0710.3104}

\bibitem[{{Minniti} et~al(1995){Minniti}, {Olszewski}, {Liebert}, {White},
  {Hill}, and {Irwin}}]{minniti+95}
{Minniti} D, {Olszewski} EW, {Liebert} J, {White} SDM, {Hill} JM, {Irwin} MJ
  (1995) {The metallicity gradient of the Galactic bulge*}. \mnras
  277:1293--1311

\bibitem[{{Nataf} et~al(2010){Nataf}, {Udalski}, {Gould}, {Fouqu{\'e}}, and
  {Stanek}}]{nataf+10}
{Nataf} DM, {Udalski} A, {Gould} A, {Fouqu{\'e}} P, {Stanek} KZ (2010) {The
  Split Red Clump of the Galactic Bulge from OGLE-III}. \apjl 721:L28--L32,
  \doi{10.1088/2041-8205/721/1/L28}, \eprint{1007.5065}

\bibitem[{{Ness} et~al(2013{\natexlab{a}}){Ness}, {Freeman}, {Athanassoula},
  {Wylie-de-Boer}, {Bland-Hawthorn}, {Asplund}, {Lewis}, {Yong}, {Lane}, and
  {Kiss}}]{ness-abu+13}
{Ness} M, {Freeman} K, {Athanassoula} E, {Wylie-de-Boer} E, {Bland-Hawthorn} J,
  {Asplund} M, {Lewis} GF, {Yong} D, {Lane} RR, {Kiss} LL (2013{\natexlab{a}})
  {ARGOS - III. Stellar populations in the Galactic bulge of the Milky Way}.
  \mnras 430:836--857, \doi{10.1093/mnras/sts629}, \eprint{1212.1540}

\bibitem[{{Ness} et~al(2013{\natexlab{b}}){Ness}, {Freeman}, {Athanassoula},
  {Wylie-de-Boer}, {Bland-Hawthorn}, {Asplund}, {Lewis}, {Yong}, {Lane},
  {Kiss}, and {Ibata}}]{ness-kine+13}
{Ness} M, {Freeman} K, {Athanassoula} E, {Wylie-de-Boer} E, {Bland-Hawthorn} J,
  {Asplund} M, {Lewis} GF, {Yong} D, {Lane} RR, {Kiss} LL, {Ibata} R
  (2013{\natexlab{b}}) {ARGOS - IV. The kinematics of the Milky Way bulge}.
  \mnras 432:2092--2103, \doi{10.1093/mnras/stt533}, \eprint{1303.6656}

\bibitem[{{Ness} et~al(2014){Ness}, {Debattista}, {Bensby}, {Feltzing}, {Ro{\v
  s}kar}, {Cole}, {Johnson}, and {Freeman}}]{ness+14}
{Ness} M, {Debattista} VP, {Bensby} T, {Feltzing} S, {Ro{\v s}kar} R, {Cole}
  DR, {Johnson} JA, {Freeman} K (2014) {Young Stars in an Old Bulge: A Natural
  Outcome of Internal Evolution in the Milky Way}. \apjl 787:L19,
  \doi{10.1088/2041-8205/787/2/L19}, \eprint{1401.0541}

\bibitem[{{Nomoto} et~al(1984){Nomoto}, {Thielemann}, and
  {Wheeler}}]{nomoto+84}
{Nomoto} K, {Thielemann} FK, {Wheeler} JC (1984) {Explosive nucleosynthesis and
  Type I supernovae}. \apjl 279:L23--L26, \doi{10.1086/184247}

\bibitem[{{Norman} et~al(1996){Norman}, {Sellwood}, and {Hasan}}]{norman+96}
{Norman} CA, {Sellwood} JA, {Hasan} H (1996) {Bar Dissolution and Bulge
  Formation: an Example of Secular Dynamical Evolution in Galaxies}. \apj
  462:114, \doi{10.1086/177133}

\bibitem[{{Nowak} et~al(2010){Nowak}, {Thomas}, {Erwin}, {Saglia}, {Bender},
  and {Davies}}]{NowThoErw10}
{Nowak} N, {Thomas} J, {Erwin} P, {Saglia} RP, {Bender} R, {Davies} RI (2010)
  {Do black hole masses scale with classical bulge luminosities only? The case
  of the two composite pseudo-bulge galaxies NGC 3368 and NGC 3489}. \mnras pp
  106--+, \doi{10.1111/j.1365-2966.2009.16167.x}, \eprint{0912.2511}

\bibitem[{{Obreja} et~al(2013){Obreja}, {Dom{\'{\i}}nguez-Tenreiro}, {Brook},
  {Mart{\'{\i}}nez-Serrano}, {Dom{\'e}nech-Moral}, {Serna}, {Moll{\'a}}, and
  {Stinson}}]{ObrDomBro13}
{Obreja} A, {Dom{\'{\i}}nguez-Tenreiro} R, {Brook} C, {Mart{\'{\i}}nez-Serrano}
  FJ, {Dom{\'e}nech-Moral} M, {Serna} A, {Moll{\'a}} M, {Stinson} G (2013) {A
  Two-phase Scenario for Bulge Assembly in {$\Lambda$}CDM Cosmologies}. \apj
  763:26, \doi{10.1088/0004-637X/763/1/26}, \eprint{1211.3906}

\bibitem[{{Ortolani} et~al(1995){Ortolani}, {Renzini}, {Gilmozzi}, {Marconi},
  {Barbuy}, {Bica}, and {Rich}}]{ortolani+95}
{Ortolani} S, {Renzini} A, {Gilmozzi} R, {Marconi} G, {Barbuy} B, {Bica} E,
  {Rich} RM (1995) {Near-coeval formation of the Galactic bulge and halo
  inferred from globular cluster ages}. \nat 377:701--704,
  \doi{10.1038/377701a0}

\bibitem[{{Phillips}(1996)}]{Phi96}
{Phillips} AC (1996) {Star Formation in Barred Galaxies}. In: {Buta} R,
  {Crocker} DA, {Elmegreen} BG (eds) IAU Colloq. 157: Barred Galaxies,
  Astronomical Society of the Pacific Conference Series, vol~91, pp 44--+

\bibitem[{{Raha} et~al(1991){Raha}, {Sellwood}, {James}, and {Kahn}}]{raha+91}
{Raha} N, {Sellwood} JA, {James} RA, {Kahn} FD (1991) {A dynamical instability
  of bars in disk galaxies}. \nat 352:411, \doi{10.1038/352411a0}

\bibitem[{{Ram{\'{\i}}rez} et~al(2000){Ram{\'{\i}}rez}, {Stephens}, {Frogel},
  and {DePoy}}]{ramirez+00}
{Ram{\'{\i}}rez} SV, {Stephens} AW, {Frogel} JA, {DePoy} DL (2000) {Metallicity
  of Red Giants in the Galactic Bulge from Near-Infrared Spectroscopy}. \aj
  120:833--844, \doi{10.1086/301466}, \eprint{astro-ph/0003116}

\bibitem[{{Rattenbury} et~al(2007){Rattenbury}, {Mao}, {Sumi}, and
  {Smith}}]{rattenbury+07}
{Rattenbury} NJ, {Mao} S, {Sumi} T, {Smith} MC (2007) {Modelling the Galactic
  bar using OGLE-II red clump giant stars}. \mnras 378:1064--1078,
  \doi{10.1111/j.1365-2966.2007.11843.x}, \eprint{0704.1614}

\bibitem[{{Rich}(1988)}]{rich+88}
{Rich} RM (1988) {Spectroscopy and abundances of 88 K giants in Baade's
  Window}. \aj 95:828--865, \doi{10.1086/114681}

\bibitem[{{Rich}(1990)}]{rich+90}
{Rich} RM (1990) {Kinematics and abundances of K giants in the nuclear bulge of
  the Galaxy}. \apj 362:604--619, \doi{10.1086/169299}

\bibitem[{{Rich} and {Origlia}(2005)}]{rich-origlia+05}
{Rich} RM, {Origlia} L (2005) {The First Detailed Abundances for M Giants in
  Baade's Window from Infrared Spectroscopy}. \apj 634:1293--1299,
  \doi{10.1086/432592}, \eprint{astro-ph/0506051}

\bibitem[{{Rich} et~al(2007){Rich}, {Origlia}, and
  {Valenti}}]{rich-origlia-valenti+07}
{Rich} RM, {Origlia} L, {Valenti} E (2007) {The First Detailed Abundances for M
  Giants in the Inner Bulge from Infrared Spectroscopy}. \apjl 665:L119--L122,
  \doi{10.1086/521440}, \eprint{0707.1855}

\bibitem[{{Rich} et~al(2012){Rich}, {Origlia}, and
  {Valenti}}]{rich-origlia-valenti+12}
{Rich} RM, {Origlia} L, {Valenti} E (2012) {Detailed Abundances for M Giants in
  Two Inner Bulge Fields from Infrared Spectroscopy}. \apj 746:59,
  \doi{10.1088/0004-637X/746/1/59}, \eprint{1112.0306}

\bibitem[{{Rojas-Arriagada} et~al(2014){Rojas-Arriagada}, {Recio-Blanco},
  {Hill}, {de Laverny}, {Schultheis}, {Babusiaux}, {Zoccali}, {Minniti},
  {Gonzalez}, {Feltzing}, {Gilmore}, {Randich}, {Vallenari}, {Alfaro},
  {Bensby}, {Bragaglia}, {Flaccomio}, {Lanzafame}, {Pancino}, {Smiljanic},
  {Bergemann}, {Costado}, {Damiani}, {Hourihane}, {Jofr{\'e}}, {Lardo},
  {Magrini}, {Maiorca}, {Morbidelli}, {Sbordone}, {Worley}, {Zaggia}, and
  {Wyse}}]{rojas-arriagada+14}
{Rojas-Arriagada} A, {Recio-Blanco} A, {Hill} V, {de Laverny} P, {Schultheis}
  M, {Babusiaux} C, {Zoccali} M, {Minniti} D, {Gonzalez} OA, {Feltzing} S,
  {Gilmore} G, {Randich} S, {Vallenari} A, {Alfaro} EJ, {Bensby} T, {Bragaglia}
  A, {Flaccomio} E, {Lanzafame} AC, {Pancino} E, {Smiljanic} R, {Bergemann} M,
  {Costado} MT, {Damiani} F, {Hourihane} A, {Jofr{\'e}} P, {Lardo} C, {Magrini}
  L, {Maiorca} E, {Morbidelli} L, {Sbordone} L, {Worley} CC, {Zaggia} S, {Wyse}
  R (2014) {The Gaia-ESO Survey: metallicity and kinematic trends in the Milky
  Way bulge}. \aap 569:A103, \doi{10.1051/0004-6361/201424121},
  \eprint{1408.4558}

\bibitem[{{Romero-G{\'o}mez} et~al(2011){Romero-G{\'o}mez}, {Athanassoula},
  {Antoja}, and {Figueras}}]{romero-gomez+11}
{Romero-G{\'o}mez} M, {Athanassoula} E, {Antoja} T, {Figueras} F (2011)
  {Modelling the inner disc of the Milky Way with manifolds - I. A first step}.
  \mnras 418:1176--1193, \doi{10.1111/j.1365-2966.2011.19569.x},
  \eprint{1108.0660}

\bibitem[{{Ryde} et~al(2010){Ryde}, {Gustafsson}, {Edvardsson}, {Mel{\'e}ndez},
  {Alves-Brito}, {Asplund}, {Barbuy}, {Hill}, {K{\"a}ufl}, {Minniti},
  {Ortolani}, {Renzini}, and {Zoccali}}]{ryde+10}
{Ryde} N, {Gustafsson} B, {Edvardsson} B, {Mel{\'e}ndez} J, {Alves-Brito} A,
  {Asplund} M, {Barbuy} B, {Hill} V, {K{\"a}ufl} HU, {Minniti} D, {Ortolani} S,
  {Renzini} A, {Zoccali} M (2010) {Chemical abundances of 11 bulge stars from
  high-resolution, near-IR spectra}. \aap 509:A20,
  \doi{10.1051/0004-6361/200912687}, \eprint{0910.0448}

\bibitem[{{Sadler} et~al(1996){Sadler}, {Rich}, and {Terndrup}}]{sadler+96}
{Sadler} EM, {Rich} RM, {Terndrup} DM (1996) {K Giants in Baade's Window. II.
  The Abundance Distribution}. \aj 112:171, \doi{10.1086/117998},
  \eprint{astro-ph/9604045}

\bibitem[{{Saha} and {Gerhard}(2013)}]{saha+13}
{Saha} K, {Gerhard} O (2013) {Secular evolution and cylindrical rotation in
  boxy/peanut bulges: impact of initially rotating classical bulges}. \mnras
  430:2039--2046, \doi{10.1093/mnras/stt029}, \eprint{1212.2317}

\bibitem[{{Saha} et~al(2012){Saha}, {Martinez-Valpuesta}, and
  {Gerhard}}]{saha+12}
{Saha} K, {Martinez-Valpuesta} I, {Gerhard} O (2012) {Spin-up of low-mass
  classical bulges in barred galaxies}. \mnras 421:333--345,
  \doi{10.1111/j.1365-2966.2011.20307.x}, \eprint{1105.5797}

\bibitem[{{Saito} et~al(2011){Saito}, {Zoccali}, {McWilliam}, {Minniti},
  {Gonzalez}, and {Hill}}]{saito+11}
{Saito} RK, {Zoccali} M, {McWilliam} A, {Minniti} D, {Gonzalez} OA, {Hill} V
  (2011) {Mapping the X-shaped Milky Way Bulge}. \aj 142:76,
  \doi{10.1088/0004-6256/142/3/76}, \eprint{1107.5360}

\bibitem[{{Samland} and {Gerhard}(2003)}]{SamGer03}
{Samland} M, {Gerhard} OE (2003) {The formation of a disk galaxy within a
  growing dark halo}. \aap 399:961--982, \doi{10.1051/0004-6361:20021842},
  \eprint{astro-ph/0301499}

\bibitem[{{Sandage}(1961)}]{San61}
{Sandage} A (1961) {The Hubble atlas of galaxies}

\bibitem[{{Schlegel} et~al(1998){Schlegel}, {Finkbeiner}, and
  {Davis}}]{schlegel+98}
{Schlegel} DJ, {Finkbeiner} DP, {Davis} M (1998) {Maps of Dust Infrared
  Emission for Use in Estimation of Reddening and Cosmic Microwave Background
  Radiation Foregrounds}. \apj 500:525--553, \doi{10.1086/305772},
  \eprint{astro-ph/9710327}

\bibitem[{{Shaw}(1987)}]{Sha87}
{Shaw} MA (1987) {The nature of 'box' and 'peanut' shaped galactic bulges}.
  \mnras 229:691--706

\bibitem[{{Shen} et~al(2010){Shen}, {Rich}, {Kormendy}, {Howard}, {De Propris},
  and {Kunder}}]{shen+10}
{Shen} J, {Rich} RM, {Kormendy} J, {Howard} CD, {De Propris} R, {Kunder} A
  (2010) {Our Milky Way as a Pure-disk Galaxy - A Challenge for Galaxy
  Formation}. \apjl 720:L72--L76, \doi{10.1088/2041-8205/720/1/L72},
  \eprint{1005.0385}

\bibitem[{{Sheth} et~al(2002){Sheth}, {Vogel}, {Regan}, {Teuben}, {Harris}, and
  {Thornley}}]{SheVogReg02}
{Sheth} K, {Vogel} SN, {Regan} MW, {Teuben} PJ, {Harris} AI, {Thornley} MD
  (2002) {Molecular Gas and Star Formation in Bars of Nearby Spiral Galaxies}.
  \aj 124:2581--2599, \doi{10.1086/343835}, \eprint{astro-ph/0208018}

\bibitem[{{Sheth} et~al(2008){Sheth}, {Elmegreen}, {Elmegreen}, and {et
  al.}}]{SheElmElm08}
{Sheth} K, {Elmegreen} DM, {Elmegreen} BG, {et al} (2008) {Evolution of the Bar
  Fraction in COSMOS: Quantifying the Assembly of the Hubble Sequence}. \apj
  675:1141--1155, \doi{10.1086/524980}, \eprint{0710.4552}

\bibitem[{{Sheth} et~al(2012){Sheth}, {Melbourne}, {Elmegreen}, {Elmegreen},
  {Athanassoula}, {Abraham}, and {Weiner}}]{SheMelElm12}
{Sheth} K, {Melbourne} J, {Elmegreen} DM, {Elmegreen} BG, {Athanassoula} E,
  {Abraham} RG, {Weiner} BJ (2012) {Hot Disks and Delayed Bar Formation}. \apj
  758:136, \doi{10.1088/0004-637X/758/2/136}, \eprint{1208.6304}

\bibitem[{{Sinha}(1979)}]{sinha+79}
{Sinha} RP (1979) {Survey of neutral hydrogen in the galactic center region}.
  \aaps 37:403--463

\bibitem[{{Smith} et~al(2004){Smith}, {Price}, and
  {Baker}}]{smith-price-baker+04}
{Smith} BJ, {Price} SD, {Baker} RI (2004) {The COBE DIRBE Point Source
  Catalog}. \apjs 154:673--704, \doi{10.1086/423248}, \eprint{astro-ph/0406177}

\bibitem[{{Soto} et~al(2007){Soto}, {Rich}, and {Kuijken}}]{soto+07}
{Soto} M, {Rich} RM, {Kuijken} K (2007) {Evidence of a Metal-rich Galactic Bar
  from the Vertex Deviation of the Velocity Ellipsoid}. \apjl 665:L31--L34,
  \doi{10.1086/521098}, \eprint{astro-ph/0611433}

\bibitem[{{Stanek} et~al(1994){Stanek}, {Mateo}, {Udalski}, {Szymanski},
  {Kaluzny}, and {Kubiak}}]{stanek+94}
{Stanek} KZ, {Mateo} M, {Udalski} A, {Szymanski} M, {Kaluzny} J, {Kubiak} M
  (1994) {Color-magnitude diagram distribution of the bulge red clump stars:
  Evidence for the galactic bar}. \apjl 429:L73--L76, \doi{10.1086/187416},
  \eprint{astro-ph/9404026}

\bibitem[{{Terndrup} et~al(1995){Terndrup}, {Sadler}, and {Rich}}]{terndrup+95}
{Terndrup} DM, {Sadler} EM, {Rich} RM (1995) {K Giants in Baade's Window. I.
  Velocity and Line-Strength Measurements}. \aj 110:1774, \doi{10.1086/117649},
  \eprint{astro-ph/9508105}

\bibitem[{{Tiede} and {Terndrup}(1997)}]{tiede-terndrup+97}
{Tiede} GP, {Terndrup} DM (1997) {A New Survey of Stellar Kinematics in the
  Central Milky Way}. \aj 113:321--334, \doi{10.1086/118255}

\bibitem[{{Tinsley}(1979)}]{tinsley+79}
{Tinsley} BM (1979) {Stellar lifetimes and abundance ratios in chemical
  evolution}. \apj 229:1046--1056, \doi{10.1086/157039}

\bibitem[{{Uttenthaler} et~al(2012){Uttenthaler}, {Schultheis}, {Nataf},
  {Robin}, {Lebzelter}, and {Chen}}]{uttenthaler+12}
{Uttenthaler} S, {Schultheis} M, {Nataf} DM, {Robin} AC, {Lebzelter} T, {Chen}
  B (2012) {Constraining the structure and formation of the Galactic bulge from
  a field in its outskirts. FLAMES-GIRAFFE spectra of about 400 red giants
  around (l, b) = (0, -10)}. \aap 546:A57, \doi{10.1051/0004-6361/201219055},
  \eprint{1206.3469}

\bibitem[{{Valenti} et~al(2013){Valenti}, {Zoccali}, {Renzini}, {Brown},
  {Gonzalez}, {Minniti}, {Debattista}, and {Mayer}}]{valenti+13}
{Valenti} E, {Zoccali} M, {Renzini} A, {Brown} TM, {Gonzalez} OA, {Minniti} D,
  {Debattista} VP, {Mayer} L (2013) {Stellar ages through the corners of the
  boxy bulge}. \aap 559:A98, \doi{10.1051/0004-6361/201321962},
  \eprint{1309.4570}

\bibitem[{{V{\'a}squez} et~al(2013){V{\'a}squez}, {Zoccali}, {Hill}, {Renzini},
  {Gonz{\'a}lez}, {Gardner}, {Debattista}, {Robin}, {Rejkuba}, {Baffico},
  {Monelli}, {Motta}, and {Minniti}}]{vasquez+13}
{V{\'a}squez} S, {Zoccali} M, {Hill} V, {Renzini} A, {Gonz{\'a}lez} OA,
  {Gardner} E, {Debattista} VP, {Robin} AC, {Rejkuba} M, {Baffico} M, {Monelli}
  M, {Motta} V, {Minniti} D (2013) {3D kinematics through the X-shaped Milky
  Way bulge}. \aap 555:A91, \doi{10.1051/0004-6361/201220222},
  \eprint{1304.6427}

\bibitem[{{Wegg} and {Gerhard}(2013)}]{wegg-gerhard+13}
{Wegg} C, {Gerhard} O (2013) {Mapping the three-dimensional density of the
  Galactic bulge with VVV red clump stars}. \mnras 435:1874--1887,
  \doi{10.1093/mnras/stt1376}, \eprint{1308.0593}

\bibitem[{{Weiland} et~al(1994){Weiland}, {Arendt}, {Berriman}, {Dwek},
  {Freudenreich}, {Hauser}, {Kelsall}, {Lisse}, {Mitra}, {Moseley}, {Odegard},
  {Silverberg}, {Sodroski}, {Spiesman}, and {Stemwedel}}]{weiland+94}
{Weiland} JL, {Arendt} RG, {Berriman} GB, {Dwek} E, {Freudenreich} HT, {Hauser}
  MG, {Kelsall} T, {Lisse} CM, {Mitra} M, {Moseley} SH, {Odegard} NP,
  {Silverberg} RF, {Sodroski} TJ, {Spiesman} WJ, {Stemwedel} SW (1994) {COBE
  diffuse infrared background experiment observations of the galactic bulge}.
  \apjl 425:L81--L84, \doi{10.1086/187315}

\bibitem[{{Williams} et~al(2011){Williams}, {Zamojski}, {Bureau}, {Kuntschner},
  {Merrifield}, {de Zeeuw}, and {Kuijken}}]{WilZamBur11}
{Williams} MJ, {Zamojski} MA, {Bureau} M, {Kuntschner} H, {Merrifield} MR, {de
  Zeeuw} PT, {Kuijken} K (2011) {The stellar kinematics and populations of boxy
  bulges: cylindrical rotation and vertical gradients}. \mnras 414:2163--2172,
  \doi{10.1111/j.1365-2966.2011.18535.x}, \eprint{1102.2438}

\bibitem[{{Woosley} and {Weaver}(1995)}]{woosley+95}
{Woosley} SE, {Weaver} TA (1995) {The Evolution and Explosion of Massive Stars.
  II. Explosive Hydrodynamics and Nucleosynthesis}. \apjs 101:181,
  \doi{10.1086/192237}

\bibitem[{{Wyse} et~al(1997){Wyse}, {Gilmore}, and {Franx}}]{WysGilFra97}
{Wyse} RFG, {Gilmore} G, {Franx} M (1997) {Galactic Bulges}. \araa 35:637--675,
  \doi{10.1146/annurev.astro.35.1.637}, \eprint{arXiv:astro-ph/9701223}

\bibitem[{{Zhao} et~al(1994){Zhao}, {Spergel}, and {Rich}}]{zhao+94}
{Zhao} H, {Spergel} DN, {Rich} RM (1994) {Signatures of bulge triaxiality from
  kinematics in Baade's window}. \aj 108:2154--2163, \doi{10.1086/117227},
  \eprint{astro-ph/9409024}

\bibitem[{{Zoccali}(2010)}]{zoccali+10}
{Zoccali} M (2010) {The Stellar Population of the Galactic Bulge}. In: {Cunha}
  K, {Spite} M, {Barbuy} B (eds) IAU Symposium, IAU Symposium, vol 265, pp
  271--278, \doi{10.1017/S1743921310000736}, \eprint{0910.5133}

\bibitem[{{Zoccali} et~al(2003){Zoccali}, {Renzini}, {Ortolani}, {Greggio},
  {Saviane}, {Cassisi}, {Rejkuba}, {Barbuy}, {Rich}, and {Bica}}]{zoccali+03}
{Zoccali} M, {Renzini} A, {Ortolani} S, {Greggio} L, {Saviane} I, {Cassisi} S,
  {Rejkuba} M, {Barbuy} B, {Rich} RM, {Bica} E (2003) {Age and metallicity
  distribution of the Galactic bulge from extensive optical and near-IR stellar
  photometry}. \aap 399:931--956, \doi{10.1051/0004-6361:20021604},
  \eprint{astro-ph/0210660}

\bibitem[{{Zoccali} et~al(2006){Zoccali}, {Lecureur}, {Barbuy}, {Hill},
  {Renzini}, {Minniti}, {Momany}, {G{\'o}mez}, and {Ortolani}}]{zoccali+06}
{Zoccali} M, {Lecureur} A, {Barbuy} B, {Hill} V, {Renzini} A, {Minniti} D,
  {Momany} Y, {G{\'o}mez} A, {Ortolani} S (2006) {Oxygen abundances in the
  Galactic bulge: evidence for fast chemical enrichment}. \aap 457:L1--L4,
  \doi{10.1051/0004-6361:20065659}, \eprint{astro-ph/0609052}

\bibitem[{{Zoccali} et~al(2008){Zoccali}, {Hill}, {Lecureur}, {Barbuy},
  {Renzini}, {Minniti}, {G{\'o}mez}, and {Ortolani}}]{zoccali+08}
{Zoccali} M, {Hill} V, {Lecureur} A, {Barbuy} B, {Renzini} A, {Minniti} D,
  {G{\'o}mez} A, {Ortolani} S (2008) {The metal content of bulge field stars
  from FLAMES-GIRAFFE spectra. I. Stellar parameters and iron abundances}. \aap
  486:177--189, \doi{10.1051/0004-6361:200809394}, \eprint{0805.1218}

\bibitem[{{Zoccali} et~al(2014){Zoccali}, {Gonzalez}, {Vasquez}, {Hill},
  {Rejkuba}, {Valenti}, {Renzini}, {Rojas-Arriagada}, {Martinez-Valpuesta},
  {Babusiaux}, {Brown}, {Minniti}, and {McWilliam}}]{zoccali+14}
{Zoccali} M, {Gonzalez} OA, {Vasquez} S, {Hill} V, {Rejkuba} M, {Valenti} E,
  {Renzini} A, {Rojas-Arriagada} A, {Martinez-Valpuesta} I, {Babusiaux} C,
  {Brown} T, {Minniti} D, {McWilliam} A (2014) {The GIRAFFE Inner Bulge Survey
  (GIBS). I. Survey description and a kinematical map of the Milky Way bulge}.
  \aap 562:A66, \doi{10.1051/0004-6361/201323120}, \eprint{1401.4878}

\end{thebibliography}
\end{document}